\documentclass[]{emulateapj}
\usepackage{apjfonts}

\usepackage{epsfig, natbib, graphicx, color, bm, verbatim, cancel, comment, amsmath}

\input epsf

\newcommand{\Mpc}{\mbox{Mpc}}

\newcommand{\bmm}[1]{\mathbf{#1}}
\newcommand{\bx}{\bmm{x}}
\newcommand{\bp}{\bmm{p}}

\newcommand{\nn}{\nonumber}

\newcommand{\avg}[1]{\left\langle #1 \right\rangle}
\newcommand{\Var}{\mbox{Var}}

\newcommand{\lkhd}{{\cal{L}}}

\newcommand{\beq}{\begin{equation}}
\newcommand{\eeq}{\end{equation}}
\newcommand{\beqa}{\begin{eqnarray}}
\newcommand{\eeqa}{\end{eqnarray}}

\newcommand{\be}{\begin{equation}}
\newcommand{\ee}{\end{equation}}

\newcommand{\bea}{\begin{eqnarray}}
\newcommand{\eea}{\end{eqnarray}}

\newcommand{\half}{\frac{1}{2}}

\newcommand{\br}{\bmm{r}}

\newcommand{\Gpc}{\mbox{Gpc}}

\newcommand{\bC}{\bmm{C}}

\newcommand{\term}{\mbox{term}}

\newcommand{\ndata}{\bar{n}}
\newcommand{\nrand}{n_R}
\newcommand{\twopt}{2pt}
\newcommand{\twoptb}{2pt_b}
\newcommand{\threept}{3pt}
\newcommand{\fourpt}{4pt}

\newcommand{\bJ}{\bmm{J}}

\shortauthors{Baxter and Rozo}
\shorttitle{A Maximum Likelihood Correlation Function Estimator}

\begin{document}
\title{A Maximum Likelihood Approach to Estimating Correlation Functions}

\author{Eric Jones Baxter}
\affil{Department of Astronomy \& Astrophysics, The University of Chicago, Chicago, IL 60637.} 
\email{ebaxter@uchicago.edu}

\author{Eduardo Rozo}
\affil{SLAC National Accelerator Laboratory, Menlo Park, CA 94025.}

\begin{abstract}
We define a Maximum Likelihood (ML for short) estimator for the
correlation function, $\xi$, that uses the same pair counting
observables ($D$, $R$, $DD$, $DR$, $RR$) as the standard \citet[][LS
  for short]{landyszalay93} estimator.  The ML estimator outperforms
the LS estimator in that it results in smaller measurement errors at
any fixed random point density.  Put another way, the ML estimator
can reach the same precision as the LS estimator with a significantly
smaller random point catalog.  Moreover, these gains are
achieved without significantly increasing the computational
requirements for estimating $\xi$.  We quantify the relative
improvement of the ML estimator over the LS estimator, and discuss the
regimes under which these improvements are most significant.  We present
a short guide on how to implement the ML estimator, and emphasize that
the code alterations required to switch from a LS to a ML estimator
are minimal.
\end{abstract}

\section{Introduction}
\label{sec:intro}

While the universe is homogeneous on large scales
\citep[e.g.][]{scrimgeouretal12}, the galaxies that populate the
universe are not distributed uniformly throughout.  Rather, galaxies
tend to cluster: we are more likely to find a galaxy in a particular
patch of the universe if that patch is near another galaxy.  The
amount of clustering is commonly characterized using the two-point
galaxy correlation function, $\xi(r)$, which can be defined as the
excess probability relative to the Poisson expectation for a galaxy to
be located in a volume element $dV$ at distance $r$ from another
galaxy:
\be
dP = \ndata\left[1 + \xi(r)\right]dV.
\ee
Here, $\bar{n}$ is the mean number density of galaxies.  

The galaxy correlation function is an extremely useful tool in
cosmology: it is relatively easy to measure using galaxy surveys
\citep[e.g.][and many
  more]{davispeebles83,hawkinsetal03,zehavietal05}, and can be used to
estimate cosmological parameters in a variety of ways \citep[for some
  recent examples
  see][]{blakeetal12,sanchezetal12,cacciatoetal12,tinkeretal12}.  With
large ongoing and near-future galaxy surveys such as the Baryon
Oscillation Spectroscopic Survey \citep[BOSS, ][]{eisenstein11} and
the Dark Energy Survey \citep[DES][]{des05}, among others, it is
increasingly important to measure the correlation function quickly and
accurately.

The most commonly used methods for determining $\xi(r)$ rely on pair
counting.  One counts the number of data-data pairs, $DD$, in the
observed galaxy catalog that have some specified radial separation, as
well as the number of random-random pairs, RR, in a randomly generated
catalog with uniform density and zero correlation.  Since $\xi$
quantifies the excess probability for two galaxies to be near each
other over the Poisson expectation, one can readily estimate the
correlation function via $\hat \xi = DD/RR-1$.  More sophisticated
estimators have been developed \citep[see][for a comparison among
  various estimators]{kerscheretal00}, with the most common estimator
in employ today being that of \citet{landyszalay93}.  The primary
advantage of pair counting techniques is that they allow complex
survey geometries and masks to be easily dealt with: one simply
applies the same mask to both the data and random catalogs when
counting pairs.

It has been shown that the correlation function estimator introduced
by \citet[][henceforth LS]{landyszalay93} is optimal (in the sense
that it has the lowest possible variance) in the limit of vanishing
correlation function and large data and random catalogs.  If these
conditions are violated --- as they are in the real world, where one
is interested in clustered fields and finite catalogs --- then it
stands to reason that the LS estimator may not be fully optimal.

In this paper we consider the Maximum Likelihood (henceforth ML)
estimator for the correlation function \citep[for a similarly minded
  but much more sophisticated approach towards estimating the power
  spectrum see][]{Jasche:2010}.  The estimator relies on the same
observables as the LS estimator --- i.e. $D$, $R$, $DD$, $RR$, and
$DR$ --- but, as we will demonstrate, it can achieve greater precision
than the LS estimator at the same number of random points.  Or
equivalently, it obtains identical precision at lower random catalog
densities, thus reducing the computation load.  We show that our
estimator reduces to the LS estimator in the limit that the
correlation function vanishes, the survey volume is very large and the
catalog densities are large, as one would expect.  Our estimator is
also very easy to implement (see the summary instructions in
\S\ref{sec:discussion}), so the effort required to switch from a
\citet{landyszalay93} estimator to that advocated in this work is
minimal.

Our work bears some similarity to an earlier analysis by
\citet{Dodelson:1997}, where they considered the full likelihood
function for a galaxy survey, i.e. the probability of finding some set
of galaxies at particular positions in a survey.  This observable
vector can, in principle, contain much more information than the pair
counts $DD$, $DR$, and $RR$, so one may expect such an analysis to be
superior to ours.  However, as noted in that work, maximizing such a
likelihood is not possible in general.  Instead, \citet{Dodelson:1997}
found that, in the limit of vanishing clustering, the maximum
likelihood estimator reduced to $\hat \xi_{ML} =(DD-DR+RR)/DD$, which
is very close to the LS estimator.  For our purposes, there are two
takeaways: first, even though \citet{Dodelson:1997} considered a much
more general problem than that of maximizing the likelihood of the
pair-counting observables, their final expression for the correlation
function only depends on pair counts in the no clustering limit.
Consequently, our analysis should not actually lose any information
relative to \citet{Dodelson:1997} in that limit.  The second takeaway
is that for clustered galaxy fields, the maximization of the
likelihood written down by \citet{Dodelson:1997} is highly
non-trivial.  As we discuss below, the maximum likelihood pair counts
estimator that we introduce easily accommodates clustering.

The layout of the paper is as follows.  In \S\ref{sec:obs} we describe
the formalism we use to define the maximum likelihood estimator for
$\xi(r)$ from the clustering observables.  In \S\ref{sec:ml_nocluster}
we apply our technique to unclustered fields, while
\S\ref{sec:ml_cluster} presents our results for clustered fields.  Our
conclusions are given in \S\ref{sec:discussion}, along with a simple
recipe for calculating the maximum likelihood correlation function
estimator.


\section{Clustering Observables and the Maximum Likelihood Estimator}
\label{sec:obs}

\subsection{Formalism and Definitions of Observables}

Let $n$ be a homogeneous random field (we will subsequently use $n$ to
refer to the number density field of galaxies, having dimensions of
$1/\rm{volume}$).  The correlation function of $n$, denoted $\xi$, can
be defined via
\bea
\xi(\br) & = & \frac{\avg{ n(\bx)n(\bx+\br) } - \avg{n(\bx)}\avg{n(\bx+\br)}}{\avg{n(\bx)}\avg{n(\bx+\br)}}  \\
	& = & \frac{\avg{ n(\bx)n(\bx+\br) } - \avg{n(\bx)}^2}{\avg{n(\bx)}^2}.
\label{eq:def}
\eea
That is, $\xi(\br)$ is simply the covariance between any two points
separated by a vector $\br$, normalized by the appropriate expectation
value.  We will further assume the field $n$ is isotropic, so that
$\xi$ depends only on the magnitude of the vector $\br$.  Our final
goal is to estimate the correlation function $\xi(r)$ of $n$ given an
empirical point realization of the field.  Specifically, given a
survey volume $V$, we assume data points within the survey are a
Poisson realization of the random field $n$.\footnote{Recent work
  by, for instance, \citet{Seljak:2009}, \citet{Hamaus:2010} and
  \citet{Baldauf:2013} has highlighted the possibility of non-Poisson
  contributions to the stochasticity of the galaxy and halo fields.  These corrections are the result of e.g. halo exclusion.
  The magnitude of such effects appears to be small (at the few
  percent level) and their inclusion in the present analysis is beyond
  the scope of this paper.  We note, however, that our framework does not preclude 
  the inclusion of such effects and future work could attempt to study how they modify the maximum 
  likelihood estimator.}

Traditional clustering estimators such as the LS estimator rely on a
set of five observables $\bx=\{D,R,DR,DD,RR\}$ from which one may
estimate the correlation function $\xi$.  For instance, the LS estimator
is given by
\be
\hat \xi_{LS} = \frac{R(R-1)}{D(D-1)}\frac{DD}{RR} - 2\frac{R-1}{D}\frac{DR}{RR} + 1,
\label{eq:xiLS}
\ee
where $D$ is the number of data points within the survey volume of
interest, and $DD$ is the number of data pairs within the radial bin
$r\pm \Delta r/2$ at which the correlation function $\xi$ is to be
estimated. $R$ and $RR$ are the corresponding quantities for a catalog
in which the positions of the data points are chosen randomly; $DR$ is
the number of data-random pairs whose separation is in the desired
range.  The form of the LS estimator presented above differs slightly
from the commonly used expression $(DD-2DR+RR)/RR$.  The additional
factors of $D$ and $R$ in Eq. \ref{eq:xiLS} allow for data and random
catalogs of different number density, while the $-1$'s correct for a
small bias in the commonly used estimator owing to the finite size of
the catalogs.

Using our model in which data points are obtained from a Poisson
random sampling of the density field $n$ we can readily compute the
expectation values and covariances (\S\ref{subsec:covariances}) of the
above observables.  We pixelize all space into pixels of volume
$\Delta V$ such that the density field $n$ is constant within a pixel.
Let $D_i$ denote the number of data points in pixel $i$, which is a
Poisson realization of the expectation value $\mu_i = n_i\Delta V$.
Since $n$ is homogeneous, the expectation value of $\mu_i$ is the same
for all pixels, with $\avg{\mu}=\ndata \Delta V$.  The probability
distribution for $D_i$ is
\be
P(D_i) = \int d\mu_i \ P(D_i|\mu_i)P(\mu_i),
\ee
where $P(D_i|\mu_i) = \exp(-\mu_i)\mu_{i}^{D_i} / D_i!$ is a Poisson distribution with mean $\mu_i$.
The first two moments of $D_i$ are
\bea
\avg{D_i} & = & \avg{\mu_i} = \ndata \Delta V \\
\avg{D_i^2} & = & \avg{\mu_i^2}+\avg{\mu_i} = (\ndata \Delta V)^2(1+\xi_0)+\ndata \Delta V.
\eea
where $\xi_0$ is the correlation function at zero separation.  

It is customary to 
recast this formalism in terms of the density fluctuation
\be
\delta_i \equiv \frac{ D_i - \avg{D_i} }{\avg{D_i}}.
\ee
By definition, $\avg{\delta_i}=0$, and 
\be
\avg{\delta_i\delta_j} = \xi_{ij} + \delta_{ij}\frac{1}{\bar n \Delta V},
\label{eq:fund}
\ee
where $\xi_{ij}=\xi(\br_{ij})$ and $\br_{ij}$ is the separation vector between pixels $i$ and $j$.
Eq. \ref{eq:fund} is the fundamental building block of our analysis.
For future reference, we note that we can rewrite $D_i$ in terms of $\delta_i$ via
\be
D_i = \bar n \Delta V(1+\delta_i).
\label{eq:Di}
\ee
Note that we have not required that $n$ be a Gaussian random field, only that it be statistically homogeneous.

We are now in a position to define our basic cluster observables.  For instance, 
the total number of data points within the survey volume is the sum
\be
D = \sum_i D_iS_i = \bar n \sum_i \Delta V (1+\delta_i)S_i,
\ee
where $S_i$ is the survey window function, such that $S_i =1$ if pixel $i$ is in the survey and $0$ otherwise.  Similarly, 
we can define the radial weighting function $W_{ij}$ such that $W_{ij} = 1$ if the pixels $i$ and $j$ are separated by
a distance $r \in [r-\Delta r/2,r+\Delta r/2]$, and $W_{ij}=0$ otherwise.  The total number of data pairs in the corresponding
radial separation bin is 
\bea
DD & = & \half \sum_{ij} D_i D_j W_{ij} S_i S_j \label{eq:dd} \\
	& = & \half \bar n^2 \sum_{ij} (\Delta V)^2 (1+2\delta_i + \delta_i\delta_j) W_{ij}S_iS_j.
\eea
The expressions for $R$, $RR$, and $DR$ are straightforward generalizations of the above formulae.


\subsection{Expectation values}
\label{subsec:expect_values}

We now turn to computing the expectation value of our observables.
The expectation value for $D$ is
\be
\avg{D} = \bar n \sum \Delta V S_i = \bar n V,
\label{eq:d_expectation}
\ee
where $V$ is the survey volume.  Likewise, the expectation value for 
$DD$ is 
\be
\avg{DD} = \half \bar n^2 \sum_{ij} (\Delta V)^2 \left(1+\delta_{ij}\frac{1}{\bar n \Delta V} + \xi_{ij} \right) W_{ij}S_i S_j.
\ee
We can zero out the Poisson term since $\delta_{ij}W_{ij}=0$.
Further, assuming the radial selection $W_{ij}$ is such that $\xi(r)$
is constant within the radial shell of interest, the above expression
reduces to
\be
\avg{DD} = \half \bar n^2 \left[ 1+\xi(r) \right] \sum_{ij} (\Delta V)^2 W_{ij} S_i S_j.
\ee
Defining the volume $V_1$ such that
\be
VV_1 = \sum_{ij} (\Delta V)^2 W_{ij} S_i S_j,
\ee
the above expression for $\avg{DD}$ can be written as\footnote{Our
  expressions are significantly simpler than those in
  \citet{landyszalay93}.  The difference is that \citet{landyszalay93}
  hold the number of points within the survey volume fixed, whereas we
  consider a Poisson sampling of a density field. This both simplifies
  the analysis, and is the more relevant problem for cosmological
  investigations.  In the limit of a large number of data points,
  however, these differences become insignificant.}
\be
\avg{DD(r)} = \half \bar n^2 V V_1 [1+\xi(r)].
\label{eq:dd_expectation}
\ee
In the limit that $r$ is much 
smaller than the survey scale, then $S_i=1$ will almost certainly
imply $S_j=1$ when $W_{ij}=1$.  Consequently, in the small scale limit,
\be
W_{ij}S_iS_j \approx W_{ij}S_i,
\label{eq:approx}
\ee
and therefore
\bea
V V_1 \approx \sum_i \Delta V S_i \sum_j \Delta V W_{ij} = VV_{shell},
\eea
where $V_{shell}$ is the volume of the shell over which the
correlation function is computed.  Note that since $W_{ij}S_i \geq
W_{ij}S_iS_j$, this approximation is in fact an upper limit,
reflecting the fact that spheres centered near a survey boundary are
not entirely contained within the survey window.  

The expectation values for the observables $R$, $RR$, and $DR$  are readily computed given the
above results.  We find
\bea
\avg{R} & = & \nrand V \label{eq:r_expectation} \\
\avg{DR} & = & \frac{1}{2}\ndata \nrand VV_1 \label{eq:dr_expectation} \\
\avg{RR} & = & \frac{1}{2} \nrand^2 VV_1. \label{eq:rr_expectation}
\eea
where $\nrand$ is the mean density of random points.  


\subsection{Covariances}
\label{subsec:covariances}

The covariance matrix between the observables can be computed in a
fashion similar to that described above \citep[for a similar approach
  going directly to $\xi$, see][]{Sanchez:2008}.  For instance,
computing the variance of $D$, we have
\bea
D^2 &=& \sum_{ij} \Delta V^2 \bar n^2 (1+\delta_i)(1+\delta_j)S_iS_j \\
&=& \sum_{ij} \Delta V^2 \bar n^2 [1+2\delta_i + \delta_i\delta_j ] S_iS_j.
\eea
Note the $\delta^0$ (first) sum reduces to $\avg{D}^2$, while the sum that is linear
in $\delta$ vanishes when we take the expectation value.  All that remains
is the $\delta_i\delta_j$ term.  Using Eq. \ref{eq:fund} we arrive at
\begin{eqnarray}
\delta_i\delta_j-\term & = & \bar n^2 \sum_{ij} \Delta V^2\left (\xi_{ij} + \frac{\delta_{ij}}{\bar n \Delta V}\right) S_i S_j  \\
 & = & \ndata V + \ndata^2 \sum_{ij} \Delta V^2 \xi_{ij} S_i S_j.
\end{eqnarray}
Putting it all together, we find
\begin{eqnarray}
\label{eq:Dvar}
\Var(D) = \ndata V + \ndata^2 \sum_{ij} \Delta V^2 \xi_{ij} S_i S_j.
\end{eqnarray}
\vspace{0.3cm}

Similar calculations can be performed for the remaining observables
and their covariances.  Appendix \ref{app:varDD} shows our derivation
of the $\Var(DD)$ as an example.  The total covariance matrix can be
expressed as a sum of a Poisson and a clustering contribution,
\be
\label{eq:cov_total}
\bC= \bC_{\mathrm{Poisson}} + \bC_{\mathrm{clustering}}.
\ee
These are
\begin{widetext}
\bea
\bC_{\mathrm{Poisson}} & = & \left( \begin{array}{ccccc}
	\ndata V & 0 & \frac{1}{2}\ndata \nrand VV_1 & \ndata^2 VV_1 & 0 \\
	\mbox{---} & \nrand V &  \frac{1}{2}\ndata \nrand VV_1 & 0 & \nrand^2 VV_1 \\
	\mbox{---} & \mbox{---} &  \frac{1}{4}\ndata \nrand V V_1 \left[ \nrand V_2 + \ndata V_2 + 1 \right] &  \frac{1}{2} \ndata^2\nrand VV_1V_2 & \frac{1}{2} \ndata\nrand^2 VV_1V_2 \\
	\mbox{---} &  \mbox{---} & \mbox{---} & \ndata^2 VV_1 \left[ (\ndata V_2)+\half \right] & 0 \\
	\mbox{---} & \mbox{---} & \mbox{---} & \mbox{---} & \nrand^2 VV_1 \left[ (\nrand V_2)+\half \right]
\label{eq:cov_poisson}
\end{array} \right) \\ & & \nn \\
\bC_{\rm{clustering}} & = & \left( \begin{array}{ccccc}
	\ndata^2 V^2 \twopt & 0 & \frac{1}{2}\ndata^2 \nrand V_{1}V^2 \twopt & \ndata^3 V_{1} V^2 \twopt + \frac{1}{2} \ndata^3 V_1V^2 \threept & 0\\
	\mbox{---} & 0 &  0 & 0 & 0 \\
	\mbox{---} & \mbox{---} &  \frac{1}{4}\ndata^2 \nrand^2 V_{1}^2 V^2 \twopt +  \frac{1}{4}\ndata^2\nrand V_1^2 V\twoptb &  \frac{1}{2} \ndata^3\nrand V_{1}^2 V^2 \twopt + \frac{1}{4} \ndata^3 \nrand V_{1}^2 V^2 \threept & 0 \\
	\mbox{---} &  \mbox{---} & \mbox{---} & \ndata^4 V_{1}^2 V^2 \twopt + \ndata^4 V_{1}^2 V^2 \threept + \frac{1}{2} \ndata^4 V_1^2 V^2\fourpt & 0 \\
	\mbox{---} & \mbox{---} & \mbox{---} & \mbox{---} & 0
\end{array} \right).
\label{eq:Cmat_clustering}
\eea
\end{widetext}
Our convention for the ordering of the observables is $\bx=\{D, R, DR, DD, RR\}$.
In the above formulae, we defined $V_2$, $2pt$, $2pt_b$, $3pt$, and $4pt$ via
\begin{eqnarray}
VV_1V_2 & = &  \sum_{ijk} \Delta V^3  W_{ij}W_{jk} S_iS_jS_k, \label{eq:V2} \\
V^2 \twopt &=& \sum_{ij} \Delta V^2  \xi_{ij}S_i S_j \label{eq:twopt} \\
(V_1^2V)\twoptb &=& \sum_{ijk} \Delta V^3 \xi_{ik} W_{ij} W_{kj} S_i S_j S_k \label{eq:twoptb} \\
(V_1 V^2) \threept &=&  \left< \sum_{ijk} \Delta V^3  \delta_i \delta_j \delta_k W_{ij} S_i S_j S_k \right> \label{eq:threept} 
\end{eqnarray}
\begin{eqnarray}
(V_1V)^2 \fourpt &=&  \sum_{ijkl} \Delta V^4  \xi_{ik} \xi_{jl} W_{ij}W_{kl} S_i S_j S_k S_l  \nonumber \\
&& + \frac{1}{2} \sum_{ijkl} \Delta V^4 C^{(4)}_{ijkl} W_{ij} W_{kl} S_i S_j S_k S_l
\label{eq:fourpt}.
\end{eqnarray}
$C^{(4)}_{ijkl}$ is the fourth order cumulant of the random field, and characterizes the non-gaussian
contribution to the 4-point term of the sample variance.

We can derive an upper limit on $V_2$ in the following way.  From
Eq. \ref{eq:V2}, we have
\begin{eqnarray}
V V_1 V_2 = \sum_{ij} \Delta V^2 W_{ij} S_i S_j \sum_k \Delta V W_{jk}S_k.
\end{eqnarray}
The sum over $k$ is less than or equal to $V_{shell}$ (regardless of the value of $j$) so that we have
\begin{eqnarray}
VV_1V_2 \leq \sum_{ij} \Delta V^2 W_{ij} S_i S_j V_{shell} = VV_1 V_{shell}.
\end{eqnarray}
Therefore, 
\begin{eqnarray}
V_2 \leq V_{shell},
\end{eqnarray}
with equality in the limit that the survey boundaries can be ignored
(i.e. if the scale of interest is very small compared to the survey
volume).  We can place a lower limit on $V_2$ using the fact that the
covariance matrix of the $R$ and $RR$ observables must be positive-semidefinite.  Enforcing this requirement yields
\begin{eqnarray}
\nrand^3V^2 V_1  \left[\nrand V_2 + \frac{1}{2}\right] - \left(\nrand^2 V V_1 \right)^2 \geq 0,
\end{eqnarray}
or
\begin{eqnarray}
\nrand V_2 \geq \nrand V_1 - \frac{1}{2}.
\end{eqnarray}
Since we can set the value of $\nrand$ to be arbitrarily large, we find
\begin{eqnarray}
\label{eq:inequality}
V_1 < V_2 < V_{shell}.
\end{eqnarray}

It is more difficult to constrain the terms in $\bC_{\rm{clustering}}$
as these depend on the details of galaxy clustering.  The $\twopt$
term can be expressed exactly as
\begin{eqnarray}
\label{eq:twopt_analytic}
\twopt = \frac{1}{(2\pi)^3}\int d^3k\ P(\vec{k}) |S(\vec{k})|^2,
\end{eqnarray}
where $P(\vec{k})$ is the galaxy power spectrum and $S(\vec{k})$ is the Fourier transform of the survey window function, i.e.
\begin{eqnarray}
S(k) = \int d^3 x\ S(\vec{x}) e^{-i \vec{k} \cdot \vec{x}}.
\end{eqnarray}

If we assume that the galaxy distribution is purely Gaussian, then the
three-point function and non-Gaussian contribution to the 4-point
functions vanish.  If we further consider the limit that the survey
volume is very large compared to the scale of interest, we can set
$W_{ij}S_i S_j \approx W_{ij} S_i$ and
\begin{eqnarray}
\label{eq:twopt_pk}
|S(k)|^2 = (2\pi)^3 V_{survey} \delta(\vec{k}).
\end{eqnarray}
In that limit, we have
\begin{eqnarray}
\label{eq:twoptb_pk}
(V_1^2 V)\twoptb &=& \frac{V_{survey}}{(2\pi)^3} \int d^3k\ P(\vec{k}) |W(\vec{k})|^2 \\
\label{eq:fourpt_pk}
(V_1V)^2\fourpt &=&  \frac{V_{survey}}{(2\pi)^3} \int d^3k\ |P(\vec{k})|^2 |W(\vec{k})|^2,
\end{eqnarray}
where $W(k)$ is the Fourier transform of the radial window function.
For a spherical survey with a step radial window function, we have
\begin{eqnarray}
\label{eq:sk}
S(k) &=& 3V_{survey} \frac{j_1(kR_{survey})}{kR_{survey}} \\
\label{eq:wk}
W(k) &=& V_{shell} j_0 (kR),
\end{eqnarray}
where $R_{survey}$ is the radius of the spherical survey and $R$ is the
radius of the scale of interest.  In addition, for the second equation we have
assumed that the shell over which the correlation function is computed
is thin.  Taking the spherical limit allows us to convert the
integrals in Eqs. \ref{eq:twopt_pk}, \ref{eq:twoptb_pk}, and
\ref{eq:fourpt_pk} into one-dimensional integrals over the power
spectrum which are straightforward to compute.


\subsection{The Landy \& Szalay Estimator as a Maximum Likelihood Estimator}

We consider now the Poisson contribution to the observable covariance
matrix written above in the limit that $\nrand\rightarrow \infty$.  In
this limit, we can think of $R$ and $RR$ as having zero variance, so
we can solve for both $V$ and $VV_1$ in terms of $R$ and $RR$:
\bea
V & = & R/\nrand \\
VV_1 & = & 2RR/\nrand^2.
\eea
Since $R$ and $RR$ are now fixed, the observable vector reduces to $\bx=\{D,DR,DD\}$, and the
corresponding covariance matrix is
\be
\bC = \left( \begin{array}{cccc}
	\bar n V & \frac{1}{2}\bar n \nrand VV_1 & \bar n^2 VV_1  \\
	 \mbox{---} &  \frac{1}{4}\bar n \nrand VV_1 \left[ (\nrand V_2) + 1 \right] &  \frac{1}{2} \bar n^2\nrand VV_1V_2 \\
	 \mbox{---} & \mbox{---} &  \bar n^2VV_1\left[ (\bar nV_2) + \half \right] \\ 
\end{array} \right),
\ee
where we have ignored the $(\bar n V_2)$ term in $\Var(DR)$ since we
are assuming $\nrand \gg \bar n$. 

Given our expressions for the means and variances of the observables
$D$, $DR$ and $DD$ and assuming a form for the likelihood function we
can evaluate the maximum likelihood estimators for $\bar n$ and $\xi$
in this limit.  We focus on $\bar n$ first.  If the only observable is
$D$, and assuming a Poisson likelihood, we arrive at 
\be
\hat n = \frac{D}{V} = \nrand \frac{D}{R}.
\ee
Using a Gaussian likelihood introduces a bias of order $1/\bar n V$
arising from the density dependence of the covariance matrix.  The
above estimator has $\avg{\hat n}=\bar n$ and $\Var(\hat n) =
\frac{\bar n}{V}$.  We can perform a similar calculation using only
the observable $DR$. Using a Gaussian likelihood and ignoring the
density dependence of the covariance matrix we arrive at
\be
\hat n = \frac{2DR}{\nrand V V_1} = \nrand \frac{DR}{RR}.
\ee

Having treated $D$ and $DR$ as independent observables, we now
consider what happens when we adopt a joint treatment.  In the limit
that the survey volume is very large compared to the scale of interest
$V_1, V_2 \rightarrow V_{shell}$ and the corresponding covariance
matrix takes the form
\be
\bC = \bar n V \left( \begin{array}{cc}
	1 & \half \nrand V_{shell} \\
	\half \nrand V_{shell} & \frac{1}{4} \nrand^2 V_{shell}^2 
	\end{array} \right).
\ee
This matrix is singular, and its zero eigenvector
is ${\bf e}=(\nrand V_{shell},-2)$.  The corresponding linear combination of observables is
\be
e= \nrand V_{shell} D - 2DR.
\ee
Its mean is $\avg{e}=0$, and since ${\bf e}$ is a zero eigenvector,
$\Var(e)=0$.  In other words, $e=0$ is a constraint equation that the
observables $D$ and $DR$ must satisfy in the large survey limit.  Note
that neither the mean nor variance of $e$ depend on $\bar n$, and
therefore the information on $\bar n$ is entirely contained in the
orthogonal eigenvector.

The orthogonal eigenvector is ${\bf e}_\perp = (2,\nrand V_{shell})$, corresponding
to an observable
\be
e_\perp = 2D + \nrand \Delta V DR.
\ee
Its mean is
\be
\avg{e_\perp} = 2\bar n V \left[ 1 + \half (\nrand V_{shell})^2 \right].
\ee
The first term in this sum stems from $D$, while the second arises
from $DR$.  In the limit that $n_R \rightarrow \infty$, $\nrand
V_{shell} \gg 1$, and therefore $e_\perp \approx DR$, so the maximum
likelihood estimator becomes that due to $DR$ alone.  If $\nrand
V_{shell} \ll 1$, one has $e_\perp \propto D$, and the joint estimator
approaches that due to $D$ alone.

We now turn to estimating $\xi$, and begin by considering the
$DR$--$DD$ observable subspace in the large survey limit.  The
corresponding covariance matrix is singular, and is given by
\be
\bC = \frac{1}{4} \bar n^3 V V_1V_2 \left( \begin{array}{cc}
	\nrand^2/\bar n^2 & 2\nrand/\bar n \\
	2\nrand/\bar n & 4 
	\end{array} \right).
\ee
The zero eigenvector is ${\bf e} = (-2,\nrand/\bar n)$, corresponding to 
\be
e = \frac{\nrand}{\bar n} DD - 2DR.
\ee
Its expectation value is $\avg{e}=\half \nrand \bar n V V_1(1+\xi)$
and again $\Var(e) = 0$.  Consequently, the equation $e=\avg{e}$ is a
constraint equation that relates $\xi$ and $\bar n$.  Explicitly, we
have
\be
\frac{\nrand}{\bar n} DD - 2DR = \half \nrand \bar n V V_1 (\xi-1).
\label{eq:constraint}
\ee

All we need to do now to find the maximum likelihood $\xi$ estimator
is to find the corresponding $\bar n$ estimator, and insert this in
our constraint equation.  To do so, we must rely on observables
orthogonal to ${\bf e}$.  Now, consider the following combination of
observables which corresponds to an orthogonal eigenvector:
\be
e_\perp = 2 DD + \frac{\nrand}{\bar n} DR,
\ee
which has an expectation value of
\be
\avg{e_\perp} = \frac{1}{2} \bar n^2 V V_1 \left[ 2(1+\xi) + \frac{\nrand^2}{\bar n^2} \right].
\ee
For $\nrand/\bar n \gg 1$, the second term dominates, and therefore
$e_\perp \approx DR$ in this limit.  The second vector orthogonal to
$e=(\nrand/\ndata)DD-2DR$ is $D$.  Thus, $D$ and $DR$ span the space
orthogonal to $e$, and therefore the relevant maximum likelihood
estimator for $\bar n$ is that discussed earlier.  For $\nrand
V_{shell} \ll 1$, the corresponding estimator is $\hat n = D/V$.
Replacing into our constraint equation for $\xi$ results in the
maximum likelihood estimator
\be
\hat \xi = \frac{R^2}{D^2} \frac{DD}{RR} - 2\frac{R}{D} \frac{DR}{RR}+1.
\ee
This is the Landy--Szalay estimator.    Conversely, if $\nrand V_{shell} \gg 1$, 
then the maximum likelihood $\hat n$ estimator is that from $DR$, $\hat n = \nrand (DR/RR)$, 
which results in
\be
\hat \xi = \frac{DD\cdot RR}{DR^2}-1.
\ee
This is the Hamilton estimator.  Both estimators are recovered in
their biased forms, but can easily be corrected to account for this
bias.  Note too that the bias scales as $1/\bar n V$, and therefore
vanishes in the limit of infinite data, as it should.

In summary, we see that in the limit that an experiment is Poisson
dominated, $n_R \rightarrow \infty$, and the survey scale is much
larger than the scale of interest, the maximum likelihood estimator
for the correlation function is either the Landy--Szalay or the Hamilton
estimator.  This suggests that neither of these estimators is optimal
for realistic surveys with finite size, finite random catalogs and/or
clustering.  It makes sense, then, to identify the true maximum
likelihood estimator to gain a lower variance estimate of the
correlation function.


\subsection{The Maximum Likelihood Estimator}

Consider the observable vector $\bx=\{D,R,DR,DD,RR\}$.  The
expectation values of the components of $\bx$ and their covariances
are given by the equations in the previous sections.  We consider
$\bp=\{\ndata, \xi, V, V_1, V_2\}$ to be unknown model parameters.
Assuming Gaussian statistics, the likelihood for the parameters $\bp$
given an observed data vector $\bx$ is
\be
\label{eq:likelihood}
\lkhd(\bp|\bx) \propto \frac{1}{\sqrt{\det \bC }} \exp\left( - \frac{1}{2}\left( \bx - \avg{\bx}
\right)^T\cdot \bC^{-1} \cdot \left( \bx - \avg{\bx} \right) \right),
\ee
where $\bC$ is the covariance matrix of the observables.  The
dependence of $\mathcal{L}$ on $\bp$ is through $\bC$ and $\avg{\bx}$.
The covariance matrix $\bC$ can be estimated from data (e.g. using
standard jackknife techniques) or from theory (e.g. with simulated
data catalogs or by developing a model for galaxy clustering).  The ML
estimator $\hat \bp_{ML}$ (which contains the ML estimator for $\xi$,
which we call $\hat{\xi}_{ML}$) is obtained by maximizing the above
likelihood with respect to the model parameters $\bp$.  There are many
routes one could take to maximize the likelihood to extract $\hat
\bp_{ML}$ (e.g. brute force, a Newton-Raphson algorithm, etc.); we
save discussion of the implementation of such methods for later.


\section{ML Performance: No Clustering}
\label{sec:ml_nocluster}

We begin by comparing the performance of the ML estimator to the LS
estimator on uniform random fields (i.e. $\xi = 0$).  As we have seen
above, for such fields in the limit that $V, \ndata V_1, \nrand V_1
\rightarrow \infty$, the LS estimator has minimal variance and is
therefore precisely the ML estimator.  Here, however, we test the
performance of the LS and ML estimators on data sets with finite
volume and point densities.

\subsection{ML Performance: Analytic Estimates}

We first wish to determine the relative performance of the LS and ML
estimators without making use of any galaxy catalogs (simulated or
otherwise).  As both LS and ML are unbiased (we checked this
explicitly), the relevant quantity for comparing the two estimators is
the error on $\xi$.  For the ML estimator, the error on
$\hat{\xi}_{ML}$ can be computed using the Fisher matrix.  For the
Gaussian likelihood we have defined in Eq. \ref{eq:likelihood}, the
Fisher matrix is given by
\be
\label{eq:fishermatrix}
F_{i j} = \frac{1}{2}\mathrm{Tr}\left[\bC_{,i} \bC^{-1} \bC_{,j} \bC^{-1} \right] 
	+ \frac{\partial \boldsymbol{\mu}^T}{\partial \bp_i} C^{-1}  \frac{\partial \boldsymbol{\mu}^T}{\partial \bp_j},
\ee
where $i$, $j$ label the components of $\bp$, and where commas
indicate partial derivatives \citep[e.g.][]{Tegmark:1997}.  The Fisher
matrix is then related to the parameter covariance matrix by
\begin{eqnarray}
\mathbf{F}^{-1} = \mathbf{C}_{param},
\end{eqnarray}
where we have used $\mathbf{C}_{param}$ to refer to the covariance
matrix of parameters to distinguish it from $\bC$, the covariance
matrix of observables.

The errors on the LS estimator can be easily
computed using propagation of uncertainty.  The variance of $\hat
\xi_{LS}$ is given by
\begin{eqnarray}
\label{eq:ls_err}
\mathrm{var}(\hat \xi_{LS}) = \bJ \bC \bJ^T
\end{eqnarray}
where the Jacobian matrix, $\bJ$, is
\begin{eqnarray}
\bJ = \left(\frac{\partial \hat \xi_{LS}}{\partial D},\frac{\partial \hat \xi_{LS}}{\partial R},  \frac{\partial \hat \xi_{LS}}{\partial DR}, \frac{\partial \hat \xi_{LS}}{\partial DD},  \frac{\partial \hat \xi_{LS}}{\partial RR}\right),
\end{eqnarray}
and where $\hat \xi_{LS}$ is given in Eq. \ref{eq:xiLS}.   
Alternatively, one can derive the above formula by expanding $\hat \xi_{LS}$
about its expectation value up to second order in fluctuations, and then evaluating
the variance of $\hat \xi_{LS}$ in a self-consistent way.


\begin{figure}[t]
\includegraphics[scale = 0.7]{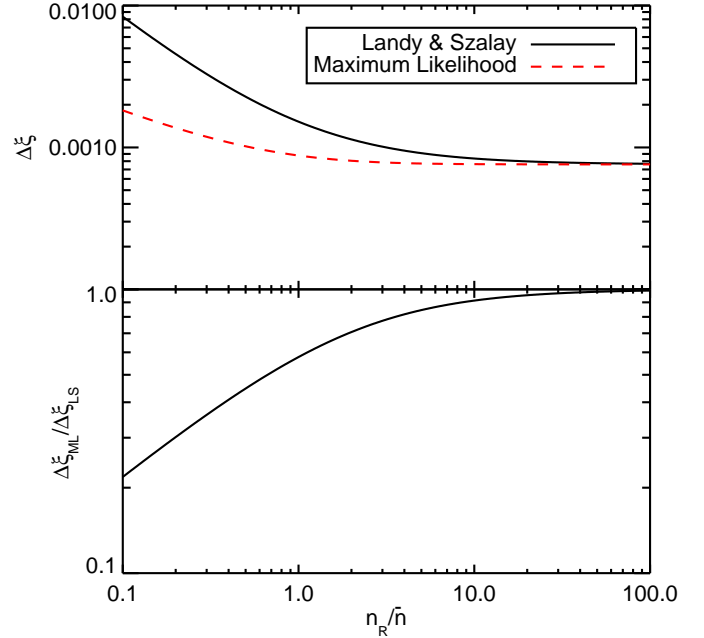}
\caption{{\it Top panel:} Standard deviation of the ML and LS
  estimators for $\xi$ as a function of the ratio of the random point
  density $\nrand$ to the data point density $\ndata$.  {\it
    Bottom panel:} The ratio of the standard deviations for the ML and LS
    estimators, $\Delta \xi_{ML}/\Delta \xi_{LS}$, as a function of
  $\nrand/\ndata$.  We have assumed a $(1~h^{-1}Gpc)^3$ survey with
  $\ndata = 5\times 10^{-5} h^3 \Mpc^{-3}$, $R = 100\ h^3 \Mpc$ and $\Delta R =
  10\ h^{-1} \Mpc$.  We have set $V_1 = V_2 = V_{shell}$.}
\label{fig:nr_converge_simple}
\end{figure}


Fig.~\ref{fig:nr_converge_simple} compares the standard deviations
$\Delta \xi_{LS}$ and $\Delta \xi_{ML}$ of the LS and ML estimators as
a function of $\nrand/\ndata$, the ratio of the number density of random
points to that of the data points.  To make this plot, we hold the
input parameter vector, $\bp_{input}$, fixed, and vary the random point
density as required.  We have chosen parameters corresponding to a
$(1\ h^{-1} \Gpc)^3$ survey with with $\ndata = 5\times 10^{-5}\ h^3
\Mpc^{-3}$, $R = 100\ h^{-1} \Mpc$ and $\Delta R = 10\ h^{-1} \Mpc$.
We have also imposed $V_1 = V_2 = V_{shell}$.  We see that both the LS
and ML estimators converge to the same value of $\Delta \xi$ at large
$\nrand$, but that the ML estimator converges much more quickly than the
LS estimator.

We now explore how this relative performance depends on the various
model parameters.  Specifically, looking back at
Eq. \ref{eq:cov_poisson}, the covariance matrix depends on four
combinations of parameters: $\ndata V$, $\nrand/\ndata$, $\nrand V_1$, and
$V_1/V_2$.  We find that varying $\ndata V$ does not affect the
relative performance of LS and ML, so we focus our attention on the
remaining three parameter combinations.  To further emphasize the
difference between the ML and LS estimators, we now focus on the
percent ``excess error'' in $\Delta\xi$ relative to the $\nrand=\infty$
value of $\Delta\xi_{ML}$, i.e. we plot
\be
\frac{\Delta \xi}{\Delta \xi_{ML}(\nrand=\infty)} - 1.
\ee
We remind the reader that in the Poisson limit that we are currently considering, both estimators yield the same $\Delta \xi(\nrand = \infty)$.

\begin{figure}[t]
\centering
\begin{minipage}{0.5\textwidth}
\centering
\includegraphics[scale=0.7]{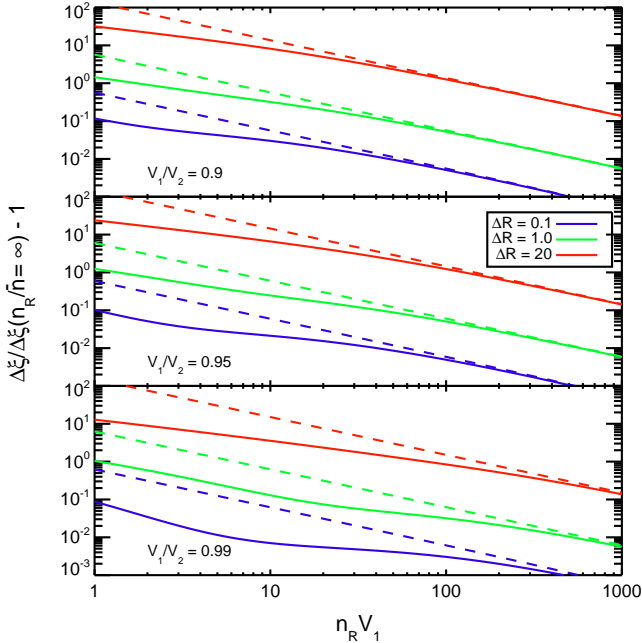}
\caption{The performance of the ML estimator relative to the LS
  estimator on uniform (Poisson) galaxy fields.  Solid lines represent the
  performance of ML while dashed lines show the performance of LS.  We
  have assume a $(1 h^{-1} \Gpc)^3$ survey and $R = 100 h^{-1} \Mpc$.  We have
  fixed $V_2 = V_{shell}$ (the most conservative assumption for the ML
  estimator) and show the effect of varying $\nrand$, $\Delta R$ and
  $V_1/V_2$. }
\label{fig:nr_converge}
\end{minipage}
\end{figure}

Our results are shown in Fig.~\ref{fig:nr_converge}.  The dashed and
solid curves show the performance of the LS and ML estimators
respectively, as a function of $\nrand$, while holding $V_1$ fixed; i.e.
the radial bin used to estimate $\xi$ is fixed.  The three sets of
curves correspond to three different values for $V_1$, or
equivalently, three different radial bin-widths.  Finally, the three
panels explore different choices of $V_1/V_2$.  Throughout, we have
set $V_2=V_{shell}$, so that varying $V_1/V_2$ is equivalent to
varying $V_1/V_{shell}$.  We expect this should provide a worst-case
scenario for the ML estimator, since the variance of $\xi$ increases
with $V_2$.


Fig. \ref{fig:nr_converge} confirms our observation that in the limit
that $\nrand$ becomes very large, the ML estimator approaches the LS
estimator.  It can also be seen in the figure that the ML estimator
becomes significantly better than LS when $\nrand V_{shell} \lesssim 100$,
and that this requirement is relatively independent of the other
parameters.  Likewise, the improvement of the ML estimator relative to the
LS estimator is stronger when $V_1/V_2 \approx 1$.

\begin{figure}
\begin{minipage}{0.5\textwidth}
\centering
\includegraphics[scale=0.7]{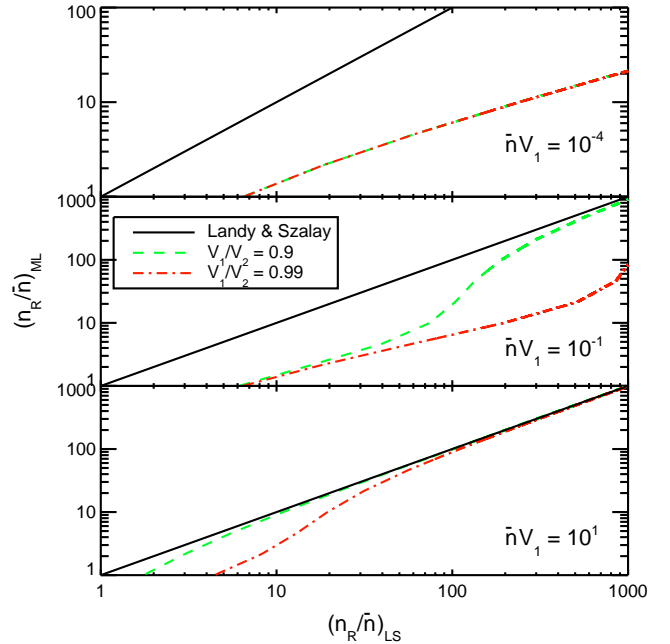}
\caption{The random point density $(\nrand/\bar n)_{ML}$ necessary for the ML 
estimator to match the precision of the LS estimator with a random point
density $(\nrand/\bar n)_{LS}$.}
\label{fig:nr_savingsfactor}
\vspace{0.1in}
\end{minipage}
\end{figure}

There is an alternative way of viewing the improved performance of the
ML estimator that is particularly well suited to a discussion of
computational efficiency.  Specifically, given an LS estimator with a
random point density $(\nrand/\ndata)_{LS}$, one can determine the
random point density $(\nrand/\ndata)_{ML}$ required for the ML
estimator to achieve the same precision.
Fig. \ref{fig:nr_savingsfactor} shows this ML random point density as
a function of the LS random point density.  Since typical pair
counting algorithms on $N$ points scale as $O(N\sqrt{N})$, this
reduction in the number of required random points means that the
computation of $\xi$ can be made significantly faster.  Because the
overall improvement depends on $V_2/V_1$, we postpone a more
quantitative discussion until after we estimate this ratio from
numerical simulations below.


As a final note before we turn our attention to numerical simulations,
we also found the
ML estimator for $\ndata$ outperforms the standard estimator
\be
\hat{\bar n} = \nrand \frac{D}{R}.
\ee
Specifically, if $D \lesssim 10^4$ and $(\nrand/n_D)\leq 10$, then the ML
estimator can outperform the standard estimator by a significant
margin.  Modern galaxy surveys have $D \gg 10^4$ galaxies, so this
result is only significant for estimating the density of rare objects,
e.g. galaxy clusters.  Note, however, that in that case, it is easy to
ensure that $(\nrand/\ndata)$ is very large, so implementing the ML
estimator is not necessary.  Still, this result was interesting enough
we thought it worth mentioning.


\subsection{ML Performance: Numerical simulation}

We have seen above that the ML estimator always performs at least as well
as LS, and that in some regimes it performs significantly better.  We now
address two related questions: 
\begin{itemize}
\item Are our Fisher matrix results representative of
  simulated data?
\item Do we expect an actual survey to fall in a regime where the ML
  estimator significantly outperforms the LS estimator?  In other
  words, what values of $V_1$ and $V_2$ are characteristic of an
  actual survey?
\end{itemize}
We address these questions through numerical simulations.

\subsubsection{Numerical Simulations}

For the $\xi = 0$ case that we are considering at this point, generating a
catalog of galaxy positions is trivial.  We focus on two possible
survey geometries:
\begin{enumerate}
\item {\bf Cube}: a cube with side length $1 h^{-1} \Gpc$
\item {\bf Survey}: a slab of dimensions $2.24~h^{-1}\Gpc$ $\times$ $2.05~h^{-1}\Gpc$
  $\times$ $0.22~h^{-1}\Gpc$\ from which we have removed a $1\ h^{-1} \Mpc$
  $\times$ 2.05~$h^{-1}$\Gpc $\times$ 0.22~$h^{-1}$\Gpc\, slice every
  10~$h^{-1}$\Mpc\, along the longest dimension of the slab.  In
    other words, this survey mask is composed of $\sim 200$ individual
    rectangular slabs separated by $1\ h^{-1} \Mpc$ gaps.
\end{enumerate}
The cube geometry is far simpler than the survey mask of any realistic survey,
while the mask adopted in the survey configuration has far more boundary effects
than any real survey is likely to have.  Thus, the combination of the two should
nicely bracket any real world scenario.

In practice, our catalog is generated in the cubical geometry and then
remapped into the slab geometry using the technique of
\cite{carlsonwhite10}.  Although this remapping procedure is
unnecessary here as the galaxies are not clustered, it will be
important when we subsequently introduce clustering (and it explains
the somewhat odd dimensions of our slab).

We perform pair counting on multiple realizations of the simulated
catalogs using a kd-tree pair counting algorithm. For illustrative
purposes, we consider two different scales:
\begin{enumerate}
\item {\bf Large}: 100-101 $h^{-1} \Mpc$ 
\item {\bf Small}: 2-3 $h^{-1} \Mpc$
\end{enumerate}
These scales are chosen to encompass the wide range over which the
correlation function is measured in actual data.  At the largest
scales, the correlation function is used as a probe of cosmology
(e.g. by measuring the BAO feature) while at the smallest scales, the
correlation function is used as a probe of galaxy formation and other
physics.

\subsubsection{Computing $V_1$ and $V_2$ on Simulated Catalogs}

It is important to accurately estimate $V_1$ and $V_2$ because, as
shown above, the effectiveness of the ML estimator depends on their
values.  As $V_2$ only enters the covariance matrix (and not the mean)
of observables, estimating it accurately requires computing the
observables over many realizations of survey volume. By contrast,
$V_1$ can be estimated easily by averaging over these realizations:
$\hat{V_1} = 2 \left< RR\right> /\left(\bar{n}^2 V \right)$, where the
angled brackets indicate an average over the different realizations.
We estimate $V_2$ by maximizing a likelihood\footnote{In practice,
  just as $V_1$ can be estimated from $RR$, $V_2$ could also be
  estimated from $RRR$, i.e. counts of triplets of random points.
  However, we have chosen not to pursue this possibility.}
\begin{eqnarray}
\label{eq:v2likelihood}
\mathcal{L}\left(V_2 | \{ \bx \} \right) &\propto& \prod_i^{N_{realizations}} \frac{1}{\sqrt{\det \bC ( V_2)}} \nonumber \\
\times &\exp& \left( - \frac{1}{2}\left( \bx_i - \avg{\bx}
\right)^T\cdot \bC^{-1}\left(V_2 \right) \cdot \left( \bx_i - \avg{\bx} \right) \right),
\end{eqnarray}
where the other parameters have been fixed. When maximizing the
likelihood, we enforce the physical requirement that $V_1 < V_2$.  

We have have used 700 realizations of the survey volume when computing
the best fit values of $V_1$ and $V_2$.  The results are summarized in
Table \ref{tab:v1v2fits}.  The first four rows of that table show the
values of $V_1/V_{shell}$ and $V_2/V_{shell}$ computed directly from
the simulations, while the final two rows show the value of $V_1/V_2$.
For the small scale case, the constraint on $V_2/V_{shell}$ is very
noisy owing to the low number of galaxies within the small scale
shells.  The noise is large enough that our best fit value of
$V_2/V_{shell}$ violates the inequality in Eq. \ref{eq:inequality}
(although it is consistent at $\sim 1.4\sigma$). Rather than use this
noisy value of $V_s/V_{shell}$ in the results that follow, we have set
$V_2 = V_{shell}$ to get the most conservative (lower) limit on
$V_1/V_2$.  For the large scale case, the noise is much less and we
are able to use the value of $V_2$ computed from the simulations.  As
we have seen above, lowering the value of $V_1/V_2$ worsens the
performance of the ML estimator.  From Table \ref{tab:v1v2fits} it is
clear that $V_1 /V_2 > 0.95$ is a conservative lower limit that should
apply even in fairly wild survey geometries.

As discussed above, the important control parameters for the ML
estimator are $\bar{n} V_1$ and $V_1/V_2$.  For a survey with a
typical number density of $\bar{n} = 5\times10^{-5} h^3 \Mpc^{-3}$, our $V_1$
results correspond to $\bar{n} V_1$ values of roughly 0.004 and 30 for
the small and large scales respectively (ignoring the relatively small
differences in $\bar{n} V_1$ for the two survey geometries).  The
value of $V_2/V_1$ is slightly more difficult to ascertain.  At the
large scale, we find that $V_1/V_2 \gtrsim 0.95$ for the three survey
geometries considered.  At small scales, our measurement of $V_2$ is
too noisy to get a good estimate of $V_1/V_2$. However, we have
demonstrated that $V_1 < V_2 < V_{shell}$ so $V_1/V_2 \geq
V_1/V_{shell} = 0.986$.  Looking back at Fig. \ref{fig:nr_converge},
we expect the ML estimator to significantly outperform the LS
estimator at small scales.  At large scales, the value of $\ndata V_1$
is large enough that we expect the improvement of LS over ML to be
more modest (although still significant for $\nrand/\ndata \lesssim
100$).  Of course, if the width of the large scale shell is reduced so
that $\nrand V_1$ goes down, we expect ML to begin to significantly
outperform LS.

\begin{table}[htpb]
\centering
\caption{Fits to $V_1$ and $V_2$ computed on numerical
  simulations.}
\label{tab:v1v2fits}
\begin{tabular}{|c|c|c|c|}
\hline
& Cube & Survey \\
\hline
\hline
$V_1/V_{shell}$ small scale & $0.9965\pm0.0008$  & $0.9862\pm0.0007$ \\
\hline
$V_1/V_{shell}$ large scale & $0.8531\pm0.0006$  & $0.6579\pm0.0003$ \\
\hline
\hline
$V_2/V_{shell}$ small scale & $1.9\pm0.9$ &  $2.3\pm0.9$ \\
\hline
$V_2/V_{shell}$ large scale & $0.886\pm0.008$ &  $0.689\pm0.005$ \\
\hline
\hline
$V_1/V_2$ small scale & $\geq 0.9965 \pm 0.0008$ &  $\geq 0.9862\pm0.0007$ \\
\hline
$V_1/V_2$ large scale & $0.963 \pm 0.009$ &  $0.955\pm0.008$ \\
\hline
\end{tabular}
\vspace{0.5cm}
\end{table}

\subsubsection{Computing the Maximum Likelihood Estimator on a Simulated Data Catalog}

The likelihood in Eq. \ref{eq:likelihood} depends on the model
parameters through both the expectation values of the observables,
$\left<\bx \right>$ and the covariance matrix $\bC$.  However, we have
found that we can obtain very accurate results by simply evaluating
the covariance matrix for parameters that are reasonably close to
$\hat \bp_{ML}$, and -- {\it keeping the covariance matrix fixed } --
vary the parameters in $\left< \bx \right>$ to maximize the
likelihood. Our approach simplifies the calculation of $\hat
\bp_{ML}$ significantly so that it reduces to several inversions of a
$5\times 5$ matrix\footnote{We will refer to the estimator calculated
  in this way as $\bp_{ML}$.  Strictly speaking, this estimator
  differs slightly from the maximum likelihood estimator of the
  previous section in that we are now fixing the covariance matrix in
  the likelihood.  The differences between the numerical values of the
  two estimators are negligible, however.}.  This means that
calculating $\hat \xi_{ML}$ is not significantly more difficult
computationally than calculating $\hat \xi_{LS}$.

We note we have explicitly verified that in the Poisson case,
the derivatives of $\bC$ can be safely neglected.
For clustered fields, this is difficult
to show in general, but one can make a rough argument.
As an illustrative example, consider the $2pt$ contribution.
In the limit of a large survey, we can rewrite Eq.~\ref{eq:twopt}
as a sum over radial bins $R'$ with $R' \leq R_{survey}$, so that
\be
V^2 2pt = V \sum V_{shell}(R') \xi(R').
\ee
Taking the derivative of $2pt$ with respect to $\xi(R)$ we find
\be
\frac{d 2pt}{d\xi(R)} = \frac{V_{shell}(R)}{V}  \ll 1,
\ee
so that, relative to the mean, the information on $\xi(R)$ from the sample variance covariance matrix is 
always being multiplied by factors of $V_{shell}(R)/V$. Perhaps from a more physical perspective,
this can also be argued by noting that the sample variance integrals are dominated by survey-volume
scale modes, with small scale modes contributing little because of the filtering by the survey window
function.

Computing the maximum likelihood estimator in the fashion described
above requires making a choice for the covariance matrix used to
analyze the data.  We will consider two possibilities for this
covariance matrix: (1) the true covariance matrix from which the data
is generated, and (2) forming an estimate of the covariance matrix
from the data itself (setting $V_2 = V_{shell}$).  The first
possibility represents the best we could hope to do: we are analyzing
the data using the same covariance matrix that was used to generate
it.

In the second case, we form an estimate of the covariance matrix from
the observed data (i.e. a new covariance matrix for each set of
observables) and then compute the maximum likelihood estimator using
this covariance matrix estimate.  To form the estimate of the
covariance matrix, we re-express the Poisson covariance matrix in
terms of clustering observables.  For instance, since $\avg{D}=\ndata
V$, and $\Var(D)=\ndata V$, we simply set $\Var(D)=D$ in the
covariance matrix.  Similarly, setting $R$, and $RR$ to their
expectation values (see \S\ref{subsec:expect_values}), we can solve
for the various terms that appear in the covariance matrix as a
function of the clustering observables $D$, $R$, and $RR$.  The
exception to this rule is $V_2$, for which we simply assume
$V_2=V_{shell}$.  The full covariance matrix obtained in this way is
\small
\be
\bC =  \left( \begin{array}{ccccc}
	D & 0 & \rho RR & 2\rho^2 RR & 0 \\
	\mbox{---} & R &  \rho RR & 0 & 2RR \\
	\mbox{---} & \mbox{---} &  \half \rho RR \left[ N_s\left(1+\rho\right) +1 \right] & \rho^2 N_s RR   & \rho N_s RR \\
	\mbox{---} &  \mbox{---} & \mbox{---} &2\rho^2 RR \left[ \rho N_s+\half \right] & 0 \\
	\mbox{---} & \mbox{---} & \mbox{---} & \mbox{---} & 2RR \left[ N_s+\half \right]
\end{array} \right),
\label{eq:approx_poisson_covmat}
\ee
%
\normalsize 
where we have defined $\rho=D/R$ and $N_s=\nrand
V_{shell}$.  Note that since $\nrand$ is known (it is chosen by the
observer), the above expression can
be computed with no a priori knowledge of the input model parameters
$\bp_{input}$. 

Given one of the above choices for the covariance matrix, we now wish
to maximize the likelihood while keeping the covariance matrix fixed.
For numerical purposes, it is convenient to reparameterize the
parameter space using a new vector $\bp' = \{\ndata,V,\alpha,\beta\} =
\{\ndata,V,VV_1,VV_1(1+\xi)\}$ such that the expectation value of the
observed data vector, $\avg{\bx}$, is linear in $V$, $\alpha$ and
$\beta$.  With this reparameterization, maximizing $\mathcal{L}$ given
$\ndata$ reduces to a simple matrix inversion problem, so the overall
minimum can be easily found using standard 1-dimensional minimization
routines.

\subsubsection{Comparing the Numerical and Analytic Calculations of the ML Estimator}

Ideally, to test the two estimators we would generate many simulated
data catalogs, perform pair counting on each one, and compute the
corresponding ML and LS estimators.  However, pair counting on many
catalogs for very high $\nrand$ is prohibitively expensive from a
computational point of view.  Instead, we make a small number of
realizations of the survey at reasonable $\nrand$ and use these
realizations to compute the unknown terms in the covariance matrix --
$V_1$ and $V_2$ -- as described above.  We then generate $10^5$ Monte
Carlo realizations of our clustering observables
$\bx=\{D,R,DR,DD,RR\}$ by drawing from a multivariate Gaussian with
the calibrated covariance matrix.  Using our Monte Carlo realizations,
we compute the mean and standard deviation of each of our estimators
in order to test whether the estimators are unbiased, and the relative
precision of the two estimators.

\begin{figure}[t]
\includegraphics[scale = 0.7]{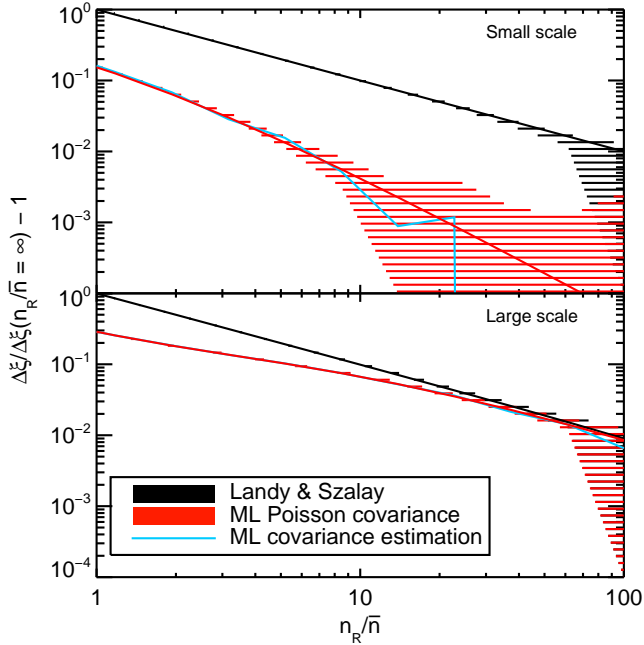}
\caption{Comparison of ML and LS estimators for $\xi$ calculated using
  simulated galaxy data in the Poisson limit.  $\Delta \xi$ is the
  standard deviation over many random realizations of the observables
  of the estimator for $\xi$.  The curves have been normalized to the
  value of the estimator for very large $\nrand/\ndata$.  The shaded
  regions represent the errors on the measured quantities owing to the
  finite number of simulations.  The solid cuves represent the
  behavior predicted from the analytic calculations described above.
  The blue curve corresponds to the ML estimator obtained when
  expressing the covariance matrix in terms of the clustering
  observables, as opposed to fixing the covariance matrix to its true
  value (and setting $V_2 = V_1$).  The survey parameters have been
  chosen to match the cubical survey geometry; upper panel is for the
  small scale, while the lower panel is for the large scale.}
\label{fig:nr_converge_simulation}
\end{figure}

Fig. \ref{fig:nr_converge_simulation} compares the result of our
numerical experiment to the analytic results presented in the last
section.  The red and black shaded regions represent the results of
our numerical experiment for the ML and LS estimators respectively;
the width of these regions represents the error on $\Delta \xi$ owing
to the finite number of realizations.  The solid red and black lines
represent the theoretical behavior predicted from the Fisher matrix as
described above.  We see that the results of our numerical experiment
are in good agreement with the results of the Fisher calculation.  For
this plot we assumed the cubical geometry discussed above; the upper
panel corresponds to the small scale, while the lower panel
corresponds to the large scale.

The blue line in Fig. \ref{fig:nr_converge_simulation} shows the ML
curve when the covariance matrix is estimated directly from the data
using the technique described above.  As can be seen, this simple
method for estimating the covariance matrix produces results that are
as good as the case when the true covariance matrix is exactly
known.\footnote{One could imagine too an iterative scheme, where the
  recovered parameters are used to re-estimate the covariance matrix.
  As shown in Figure \ref{fig:nr_converge_simulation}, however, this
  is not necessary.}  Thus, we can firmly conclude that in the regimes
described above -- namely low $\nrand V_1$ and $V_1/V_2 \sim 1$ -- the ML
estimator represents a significantly more powerful tool for estimating
the correlation function than the LS estimator.


\section{ML Performance: Clustered Fields}
\label{sec:ml_cluster}

Clustering of galaxies introduces new terms into the covariance matrix
of the observables $\bx=\{D,R,DR,DD,RR\}$.  These new terms involve
various integrals of the correlation function and its higher moments
over the survey volume (see Eq. \ref{eq:Cmat_clustering}).
Consequently, we no longer expect the LS estimator to be the large
volume, large $\nrand$ limit of the ML estimator.

Our program here will be very similar to that discussed above for the
case with no clustering.  The major difference is that in the present
case, the covariance matrix of the observables is more difficult
to calculate as it depends on the detailed clustering properties of
the galaxies.  One consequence of this fact is that we cannot easily
include the dependence of the covariance matrix on $\xi$ in the Fisher
matrix calculation as we did previously.  Instead, we will consider
the covariance matrix to be fixed.  This means that we are throwing
out some information, but as we have seen above, including the
dependence of the covariance matrix on the model parameters does not
significantly affect our constraints on $\xi$.  

\subsection{ML Performance: Analytic Estimates}

As we did previously, we can estimate the errors on our ML estimate of
$\xi$ using the Fisher matrix.  We set $C_{,i}=0$ in
Eq. \ref{eq:fishermatrix} as the dependence of $C$ on $\xi$ is not
known.  To estimate $C$ requires computing the $\twopt$, $\twoptb$,
$\threept$, and $\fourpt$ terms.  For our Fisher analysis, we choose
to perform the calculation of these terms analytically assuming the
Gaussian, large volume, spherical survey limit discussed above.  This allows
us to express the clustering terms as one-dimensional integrals over
the power spectrum (i.e.  Eqs. \ref{eq:twopt_pk}, \ref{eq:twoptb_pk},
\ref{eq:fourpt_pk}, \ref{eq:sk}, and \ref{eq:wk}).  Therefore, given a
power spectrum, we can compute the full covariance of the observables,
which in turn allows us to compute the Fisher matrix, and therefore,
the error on $\xi$.  We use the power spectrum output from CAMB
\citep{Lewis:1999} assuming standard $\Lambda$CDM cosmological
parameters that are consistent with the results of WMAP9
\citep{Bennett:2012}: $h= 0.7$, $\Omega_{c}h^2 = 0.1127$, $\Omega_b
h^2 = 0.02254$, $n_S = 1.0$.  To convert the matter power spectrum
into a galaxy power spectrum we have assumed a constant bias of $b=2$,
appropriate for galaxies in a BOSS-like sample.

\begin{figure}[t]
\begin{center}
\includegraphics[scale = 0.7]{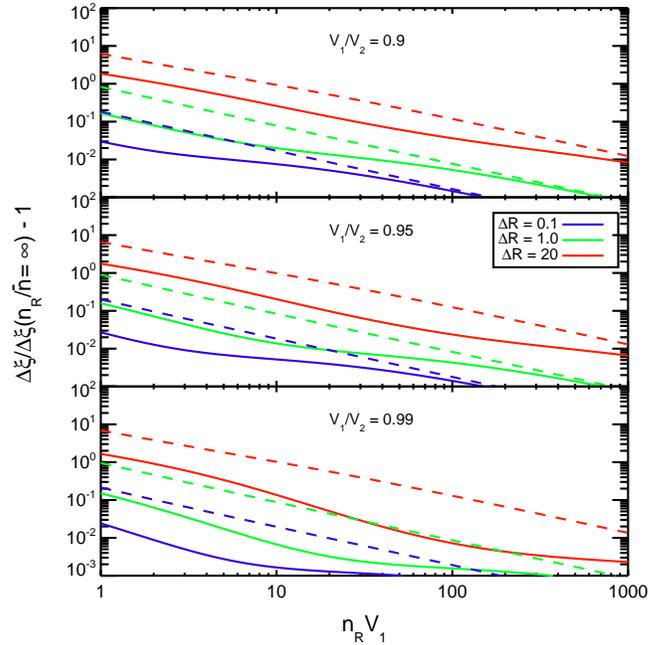}
\caption{The performance of the ML estimator relative to the LS
  estimator on clustered galaxy fields (analogous to
  Fig. \ref{fig:nr_converge} for unclustered fields).  See text for
  details of our assumptions about cosmology and bias.  Solid lines
  represent the performance of ML while dashed lines show the
  performance of LS.  The errors on ML have been computed using the
  Fisher matrix, assuming a spherical survey with a volume of 1000
  $h^3 \Mpc^3$, $\ndata = 5\times10^{-5} h^3 \Mpc^{-3}$, and $R = 100
  h^{-1} \Mpc$.  We have set $V_1 = 0.97 V_{shell}$, a resonable value
  for an actual survey (as illustrated above).  We have set $V_2 =
  V_{shell}$, which gives us the most conservative limit on the
  performance of the ML estimator.}
\label{fig:nr_converge_clustering_theory}
\end{center}
\end{figure}

Fig. \ref{fig:nr_converge_clustering_theory} shows the results of the
Fisher analysis including the effects of galaxy clustering.  This
figure is analogous to the earlier Fig. \ref{fig:nr_converge} which
applied to unclustered fields.  Comparing the two figures reveals that
clustering enhances the performance of ML relative to LS somewhat but
that otherwise the qualitative behavior is very similar.  The results
from our discussion of the ML estimator on unclustered fields
therefore carry over to clustered fields mostly unchanged.


\subsection{ML Performance: Numerical Simulation}

As we have done for the unclustered fields, we would now like to connect
the results of our Fisher matrix study to results obtained from
analyzing simulated realizations of the pair counts observables for
clustered galaxy fields.  There are two reasons for going beyond the
Fisher marix estimates. First, the simulated realizations allows us to
measure $\bC_{\rm{clustering}}$ in a survey that is more realistic
than a sphere; as we will see, this change in geometry significantly
impacts the values of the clustering terms.  Second, having simulated
data allows us to experiment with analyzing this data using different
covariance matrices.  This is important, as estimating the true
observable covariance matrix for clustered galaxy fields is
non-trivial.

Measuring the clustering contribution to the covariance matrix
requires realizations of a clustered galaxy field.  While it is
conceivable that the observable covariance matrix could be estimated
from a single cosmological realization using a jackknife, we have
found that this approach does not yield reliable results at large
scales.  In the jackknife approach, chunks of the survey volume ---
which must be significantly larger than the scales of interest --- are
removed and the pair observables are recalculated; the covariance
matrix of the observables can then be related to the covariance across
the jackknives.  We suspect that the reason this approach does not
work in practice is that when dealing with large scales, the size of
the removed chunks becomes significant enough that they effectively
change the values of $V$, $V_1$, and $V_2$ relative to what they were
before each chunk was removed.  Consequently, the pair observables
computed on the jackknifed survey volume are not drawn from the same
underlying covariance matrix as the observables in the full survey
volume.

Rather than attempt to address the problems with the jackknives, we
instead estimate the observable covariance matrix from multiple
realizations of an N-body simulation.  We use 41 cosmological
realizations of a $(2400 \,h^{-1}\Mpc)^3$ volume produced by the
LasDamas group \citep{McBride:2011}.  The LasDamas simulations assume
a flat $\Lambda$CDM cosmology described by $\Omega_m = 0.25$,
$\Omega_{\Lambda} = 0.75$, $\Omega_b = 0.04$, $h = 0.7$, $\sigma_8 =
0.8$ and $n_s = 1$.  For our ``galaxies'', we rely on the halo
catalog, randomly selecting halos (restricting to $M \geq 10^{13}\,
h^{-1} M_{\odot}$) to achieve the desired number density.  Henceforth,
we will consider the measurement of the correlation function in a
radial bin extending from $R = 50\,h^{-1}\Mpc$ to $R= 60\,h^{-1}\Mpc$.
This radial bin was chosen as it is small enough to be significantly
impacted by clustering and big enough to contain a large number of
galaxies so that the impact of counting noise is minimized.

\begin{figure}[t]
\begin{center}
\includegraphics[scale = 0.7]{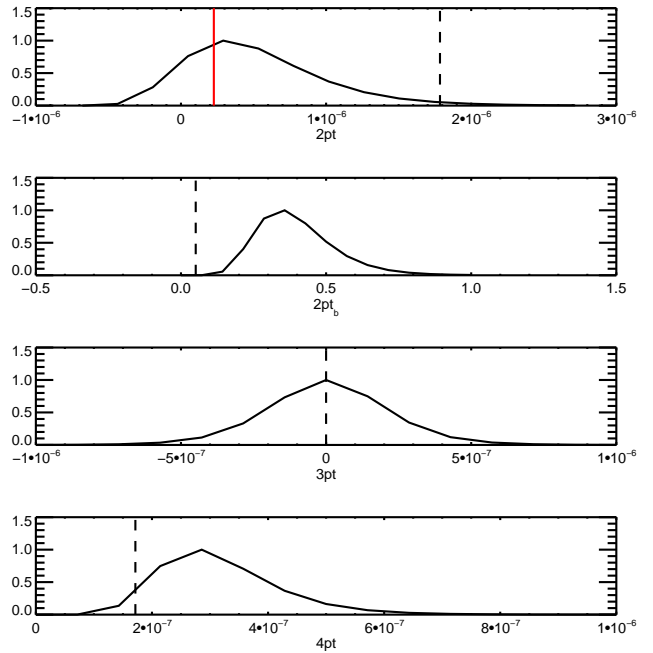}
\caption{The results of our fit for the clustering terms in the
  covariance matrix to data from N-body simulations.  The curve in
  each panel represents the probability distribution from the fit for
  the corresponding parameter.  The solid vertical line (red in the
  online version) in the top panel represents our prediction for
  $\twopt$ from integrating Eq. \ref{eq:twopt_analytic} assuming the
  true cubical geometry of the simulation.  Dashed vertical lines in
  all panels represent predictions for the n-point terms obtained by
  integrating the expressions we have derived for them assuming the
  large volume, spherical survey limit (see \S\ref{subsec:covariances}).}
\label{fig:npoint_fits}
\end{center}
\end{figure}

We estimate the $\twopt$, $\twoptb$, $\threept$, and $\fourpt$ terms
in a manner similar to that used to estimate $V_2$ above.  First, we
estimate $\ndata$, $\nrand$, $V$ and $V_1$ using the expressions we have
derived for the expectation values of the pair counts observables.
$V_2$ is then estimated by maximizing a likelihood as in
Eq. \ref{eq:v2likelihood}.  Since the $D$, $DR$ and $DD$ observables
are now affected by clustering, however, we consider only the $R$ and
$RR$ observables when evaluating the likelihood in
Eq. \ref{eq:v2likelihood}.  With estimates of $\ndata$, $\nrand$, $V$, $V_1$ and
$V_2$ in hand, we can form an estimate of $\bC_{\rm{Poisson}}$.
Finally, to determine the clustering terms, we maximize the four
dimensional likelihood function defined by
\begin{align}
& \mathcal{L}\left(\{\twopt, \twoptb, \threept, \fourpt \}  | \{ \bx \} \right)  \nonumber \propto \nonumber \\
& \prod_{i=1}^{N_{realizations}} \frac{1}{\sqrt{\det \bC }} \times \exp \left( - \frac{1}{2}\left( \bx_i - \avg{\bx} \right)^T\cdot \bC^{-1} \cdot \left( \bx_i - \avg{\bx} \right) \right) 
\end{align}
where
\begin{eqnarray}
\bC = \bC_{Poisson} + \bC_{clustering}(\twopt, \twoptb, \threept, \fourpt),
\end{eqnarray}
and, in our case, $N_{realizations} = 41$.  The results of this fitting procedure
are shown in Fig. \ref{fig:npoint_fits}.  It is clear from the figure
that we obtain no significant detection of the $\twopt$ and $\threept$
terms.  The fact that there is no signficant $\threept$ detection is
not surprising as the galaxy field is roughly Gaussian.  The
non-detection of the $\twopt$ term, on the other hand, can be
attributed to the fact that we only have $41$ realizations of the
survey volume and therefore our $\twopt$ estimate is noisy.  The numerical
results for our fits are shown in Table \ref{tab:npoint_fits}.

Although we apparently do not have the constraining power to robustly
estimate the $\twopt$ term from the N-body simulations, we can
estimate it analytically using Eq. \ref{eq:twopt_analytic}.  For the
cubical geometry that we consider here, we can integrate
Eq. \ref{eq:twopt_analytic} exactly to obtain an estimate for
$\twopt$.  Performing this calculation requires an estimate of the
power spectrum, $P(k)$, which we obtain from CAMB \citep{Lewis:1999}.
The bias is determined by matching the prediction for $\xi$ from the
power spectrum to its measured value at the scale of interest; we find
that the bias is roughly $b = 1.6$.  Our analytic estimate of $\twopt$
is shown as a red line in Fig. \ref{fig:npoint_fits}; the numerical
value of $\twopt = 2.2\times10^{-7}$ is in good agreement with the
result from the N-body simulations shown in Table
\ref{tab:npoint_fits}.


\begin{table}[htpb]
\centering
\caption{Fits to clustering terms computed from N-body simulations.}
\label{tab:npoint_fits}
\begin{tabular}{|c|c|c|}
\hline
& $R = 50 h^{-1} \Mpc$, $\Delta R = 10 h^{-1}\Mpc$ \\
\hline
\hline
$\twopt$  & $(4\pm4)\times 10^{-7}$ \\
\hline
$\twoptb$ & $0.4\pm0.1$ \\
\hline
$\threept$ & $(0 \pm 6) \times 10^{-8}$ \\
\hline
$\fourpt$ & $(3 \pm 1) \times 10^{-7}$\\
\hline
\end{tabular}
\vspace{0.5cm}
\end{table}

\begin{figure}[t]
\begin{center}
\includegraphics[scale = 0.7]{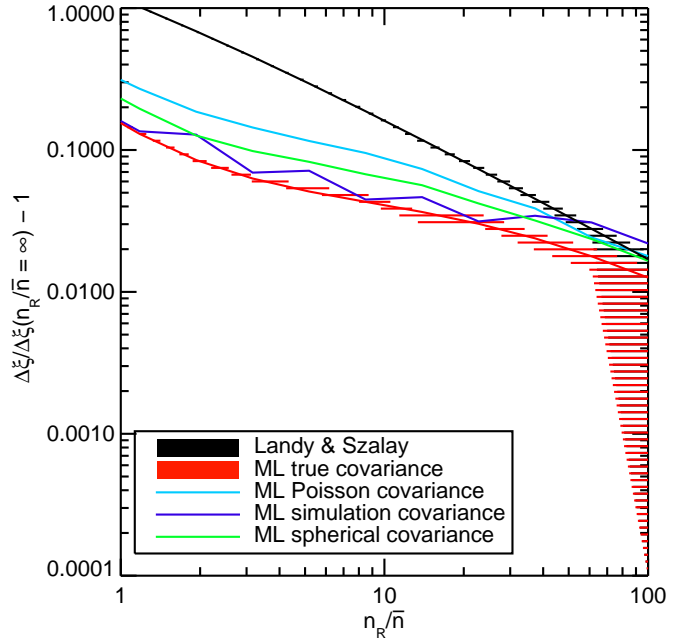}
\caption{The performance of the ML estimator relative to the LS
  estimator on clustered galaxy fields (analogous to
  Fig. \ref{fig:nr_converge_simulation} for unclustered fields). Black
  and red lines represent analytic predictions for the LS (using
  Eq. \ref{eq:ls_err}) and ML estimator (using the Fisher matrix)
  respectively.  Black and red shaded regions represent the error
  bands corresponding to these two estimators computed from numerical
  simulations.  The remaining curves represent the results obtained
  when the simulated data is analyzed with the ML estimator using
  different choices of the covariance matrix.  Blue corresponds to
  using a Poisson covariance matrix estimated from the data; purple
  corresponds to using the mean covariance matrix computed from $40$
  realizations of the survey volume; green corresponds to computing
  the clustering contribution covariance matrix analytically as
  described in the text (using the spherical approximation for the
  $\twoptb$ and $\fourpt$ terms).}
\label{fig:nr_converge_clustering_numerical}
\end{center}
\end{figure}

With our estimates of the $\twoptb$, $\threept$, and $\fourpt$ terms
from the fits to the cosmological realizations, and our estimate of
$\twopt$ by direct integration, we can now compute the full covariance
matrix of observables.  We use this covariance matrix to generate
realizations of the observables as we have done above for the case
without galaxy clustering. The results of our analysis of these
simulated data sets are presented in
Fig. \ref{fig:nr_converge_clustering_numerical}.  As in
Fig. \ref{fig:nr_converge_simulation}, the black curve represents the
analytic prediction for the errors obtained using the LS estimator and
the shaded region represents the measurement of the errors on
simulated data (the width of the region shows the error on this
measurement owing to a finite number of realizations).  The red curve
represents the prediction from the Fisher matrix for the ML estimator,
while the red shaded region represents the numerical results computed
by analyzing the data using the covariance matrix that was used to
generate it.

The analyst wishing to compute the correlation function in a galaxy
survey with the ML estimator must first estimate the covariance matrix
of $D$, $DR$, $DD$, and $RR$.  In the unclustered case considered
previously, the estimation process was straightforward: we simply computed
$\bar{n}$, $V$, and $V_1$ from the observables $D$, $R$, and $RR$, 
and we set $V_2 = V_1$, and then
substituted into Eq. \ref{eq:cov_poisson}.  We showed (blue curve in
Fig. \ref{fig:nr_converge_simulation}) that this procedure works well.
In the present case, however, computing the observable covariance
matrix is more difficult as it depends on the $\twopt$, $\threept$,
$\fourpt$, and $\twoptb$ terms, which are not known a priori, and are
difficult to estimate from the data.

There are several ways around this difficulty.  The simplest is to
ignore the clustering contribution to the covariance matrix and simply
compute the covariance matrix in the Poisson limit as above (using
only the $R$ and $RR$ observables as these are unaffected by galaxy
clustering) for the purposes of defining the ML estimator.  Figure
\ref{fig:nr_converge_clustering_numerical} compares the performance of
this simple approach (blue curve) to the LS estimator and to the ML
estimator when run using the correct covariance matrix. It appears
that the Poisson covariance matrix approach generally does better than
LS but that it does not achieve the maximal performance that can be
obtained with the ML estimator using the true covariance matrix.  We
note that we have also checked that using the Poisson covariance
matrix did not bias the resulting ML estimator.

Alternatively, one can use our analytic estimates of the covariance matrix
of clustered fields to analyze the data.  There are
several ways that one could go about this in deail; we take an approach
that requires little computational work.  As the $\twopt$ function can
be easily estimated by integrating Eq. \ref{eq:twopt_analytic} for a
cubical geometry, we estimate $\twopt$ in that way.  The remaining
clustering terms ($\twoptb$, $\threept$, and $\fourpt$) are more
difficult to estimate in a cubical geomtry, but can easily be
estimated for a spherical geometry using Eqs. \ref{eq:twoptb_pk} and
\ref{eq:fourpt_pk} (we assume that the distribution is Guassian so
that the $\threept$ term vanishes).  As seen in
Fig. \ref{fig:npoint_fits}, these estimates of the clustering terms
are not perfect, but they at least give us some handle on the
magnitude of the clustering contribution.  Our easy-to-compute
estimate of the clustering contribution to the covariance matrix can
then be combined with an estimate of the Poisson contribution (as above)
to form an estimate of the total covariance matrix.  The green curve
in Fig. \ref{fig:nr_converge_clustering_numerical} shows the results
of analyzing the data using this covariance matrix.  We see that this
approach generally does better than using only the Poisson covariance
matrix, but that it does not do quite as well as using the true
covariance matrix.

Finally, one could derive accurate estimates by running several
numerical simulations and computing the covariance of the observables
across these simulations.  Such simulations are typically already
perfomed in order to estimate statistical uncertainties, and so this
step should not require any additional overhead.  This approach can be
simulated by using some small number, $N_{realization}$, of the
realizations to compute an observable covariance matrix, and then
analyzing the data using this covariance matrix.  We test this method
by using our input covariance matrix to generate $N_{realizations}=40$
independent data realizations, which in turn are used to estimate the
covariance matrix of the generated data.  This estimated covariance
matrix is then used to estimate $\xi$.  The purple curve in a new
independent data set.  Fig. \ref{fig:nr_converge_clustering_numerical}
shows the errors on $\xi$ calculated in this way.  We see
that 40 realizations is sufficient to capture the clustering
information in the covariance matrix for the purposes of the ML
estimator.

One might worry about the circularity of estimating the covariance
matrix in the manner described above; after all, the clustering terms
in Eq. \ref{eq:Cmat_clustering} depend precisely on the correlation
function that we are trying to measure.  However, as we have seen
above, incorrectly estimating the covariance matrix does not lead to a
bias in the recovered $\hat{\xi}_{ML}$, but rather increases its
variance.  Furthermore, even if the clustering terms are set to zero,
we still get a lower variance estimate of $\xi$ than with LS.  Any
reasonable errors in the cosmology used to estimate the covariance
matrix will never cause the ML estimator to perform worse than LS.
Finally, if an analyst is really worried about obtaining the
absolutely minimum variance estimator of $\xi$, it is always possible
to apply the ML estimator in an iterative fashion.  One simply assumes
a cosmology when calculating the observable covariance matrix, and
then adjusts the cosmology based on the recovered $\hat{\xi}_{ML}$;
the process can then be repeated until convergence is reached.


\section{Discussion}
\label{sec:discussion}

We have explored the utility of the ML estimator for computing galaxy
correlation functions.  The ML estimator makes use of the same
pair-counting observables as the standard LS estimator, yet the former
significantly outperforms the latter in certain regimes.  Moreover,
because all but one of the parameters in the likelihood model are
linear (assuming we use a fixed covariance matrix, as described
above), the likelihood maximization is numerically trivial.
Consequently, we see no reason not to switch from LS estimators to ML
estimators: the ML estimator is always better, and has no significant
computational requirements in excess of those for the LS estimator.
In short, there are only upsides to using the ML estimators, and no
real downsides.


\begin{figure}[t]
\begin{center}
\includegraphics[scale = 0.7]{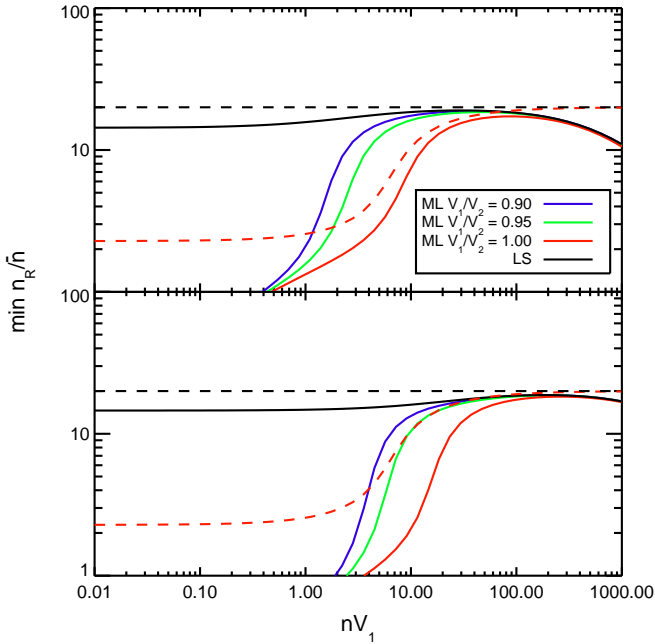}
\caption{The minimum value of $\nrand$ required to achieve convergence at
  the 5\% level to the value of $\Delta \xi$ at $\nrand = \infty$ , as a
  function of $\ndata V_1$.  We have assumed a spherical survey with
  volume $V = (2400 h^{-1} \Mpc)^3$.  The two panels correspond to
  different radial scales: upper panel is for $R = 50 h^{-1}\Mpc $ to
  $60h^{-2}\Mpc$ while lower panel is for $R = 100 h^{-1}\Mpc $ to
  $101h^{-2}\Mpc$.  Dashed lines (which do not depend on the radial
  scale) represent the case where there is no clustering.}
\label{fig:nr_converge_minnr}
\end{center}
\end{figure}


For an analyst wishing to compute the correlation function from a
galaxy survey, an important question is how large must the random
catalog be in order to get errors on $\xi$ that are close to what
would be obtained with an infinite random catalog?  In
Fig. \ref{fig:nr_converge_minnr} we plot the minimum value of
$\nrand/\ndata$ required to obtain errors on $\xi$ that are within 5\% of
the value of $\Delta \xi$ at $\nrand = \infty$ for both the ML and LS
estimators.  As we have discussed previously, the ML estimator allows
one to compute the correlation function to the same precision as with
LS while using a significantly smaller random catalog.  We see that LS
acheives convergence to the 5\% level at $\nrand/\ndata \sim 15$.
Depending on the value of $\ndata V_1$, the ML estimator can reduce
the required value of $\nrand$ by up to a factor of 7.

Perhaps the single biggest obstacle one faces from the point of view
of implementing the ML estimator is that one must specify the
covariance matrix used to minimize $\chi^2$.  In practice, however, we
do not believe this is a particularly problematic issue.  Firstly,
modern cosmological analysis typically rely on extensive numerical
simulations to calculate the covariance matrix of survey observables.
Just as one can use these numerical simulations to calibrate the
variance in $\hat \xi_{LS}$, one can use the same simulations to
estimate the covariance matrix of the observables $D$, $R$, $DR$,
$DD$, and $RR$.  With this covariance matrix at hand, one can then
compute $\hat \xi_{ML}$, and use these same simulations to estimate
the error $\Delta\hat \xi_{ML}$.  Alternatively, because the
simulation-based covariance matrices are expected to be correct, one
could, if desired, treat the problem using Bayesian statistics as
usual, without fear of underestimating uncertainties.

Our work can be compared to a recent paper by \citet{Vargas-Maga:2012}
(hereafter VM).  In that work, the authors construct an estimator that
significantly outperforms LS on realistic surveys.  In brief, their
estimator is a linear combination of all possible ratios of the
standard pair counts observables $\bx=\{DD, DR, RR\}$ up to second
order (see their Table 1), i.e.
\be
\hat \xi = c_0 + \sum_{i=1}^{6} c_i R_i + \sum_{y}^{18} c_iR_i^{(2)},
\ee
where $R_i$ are various ratios of elements of $\bx$.
The set of coefficients is calibrated using lognormal simulations
of the survey with a known correlation function,
with the coefficients dependent on the simulations and the survey
geometry.  

At first sight, an obvious objection to this approach is that because
the optimal coefficients depend on the correlation function of the
field, the sensitivity of the resulting estimator to choice of
correlation function in the log-normal simulations makes this method
undesirable.  However, VM demonstrated that this problem can be solved
with an iterative technique: one uses LS to estimate $\xi$, and then
uses that $\xi$ to generate log-normal realizations, so that the data
itself informs the simulations.  These realizations are used to define
the coefficients, which are then employed in the data to get a new
estimate of $\xi$, and the whole procedure is iterated until
convergence is achieved.

When we run our analysis mirroring the random point densities and
binning scheme of VM, we find that the improvements of the ML
estimator relative to the LS estimator are on the order of several
percent, significantly lower than the improvements advertised by VM.
Note, however, that Figure \ref{fig:nr_converge_clustering_numerical}
demonstrates that the ML estimator saturates the Cramer--Rao bound on
the variance of $\hat \xi$.  As the estimator of VM uses the same pair
counts observables as the ML estimator, the fact that the former
violates the Cramer--Rao bound may at first appear problematic.  The
resolution to this problem is that for an arbitrary correlation
function, the procedure of VM is biased and therefore the Cramer--Rao
bound does not apply.  The VM estimator is only unbiased for
correlation functions whose shape can be fit by the particular form
assumed in their iterative fitting procedure (Eq. 3 of their paper).

As a summary, we would say that if one wishes to quickly and easily
estimate an arbitrary correlation function, we can unambiguously
advocate the use of the ML estimator over the LS estimator.  Under
some circumstances, however, where the correlation function is known
to be well fit by the form assumed by VM, their iterative scheme leads
to a dramatic reduction of errors, at the expense of increased
computational requirements and complexity.

\subsection{Recipe for Computing the Maximum Likelihood Correlation Function Estimator}

To aid the reader, we now provide a step-by-step guide of the steps
required to implement our ML estimator.  

\begin{enumerate}
\item Compute the observables $[D,R,DR,DD,RR]$ in the usual fashion.
  \item Estimate the covariance matrix of observables:
  \begin{itemize}
    \item In most cases, we expect this to be done via numerical simulations.
    \item One may use a Poisson covariance matrix to analyze the data
      as in Eq. \ref{eq:approx_poisson_covmat}.
    \item If desired/necessary, add analytic estimates of the
      clustering terms to the covariance matrix.
  \end{itemize}
\item Maximize the likelihood defined in Eq. \ref{eq:likelihood} to
  find $\hat \xi_{ML}$, keeping the covariance matrix fixed to the
  estimate from above.  To do so, use the parameter vector
  $\bp'=\{\ndata,V,\alpha,\beta\}=\{\ndata,V,V V_1,V V_1(1+\xi)\}$,
  and minimize $\chi^2$.  With this redefinition, the only non-linear
  parameter in our expressions for the expectation values of the
  observables (Eqs. \ref{eq:d_expectation}, \ref{eq:dd_expectation},
  \ref{eq:r_expectation}, \ref{eq:dr_expectation},
  \ref{eq:rr_expectation}) is $\ndata$.  Consequently, minimization
  can easily be achieved be defining a grid in $\ndata$.  For each
  grid point, one finds the maximum likelihood value for the linear
  parameters through straightforward matrix inversion, and then
  evaluates the likelihood.  The overall minimum can easily be
  estimated from the data grid.
\end{enumerate}

Before we end, there in one last additional point that is worth noting
with regards to correlation function estimators.  In particular, our
formalism and maximum likelihood framework also suggests what are
ideal binning conditions.  Specifically, in order to gaurantee that
the Gaussian likelihood approximation is good, one should adopt radial
bins such that $DD\gg 1$.  If one sets $DD=100$, the corresponding bin
width $\Delta \ln R = \Delta R/R$ for a scale R should be
\be
\Delta \ln R = 1.59 \left( \frac{1\ \Mpc}{R} \right)^3 \left( \frac{1\ \Gpc^3}{V} \right) \left( \frac{10^{-4}\ \Mpc^{-3}}{\bar n} \right)^2 
\label{eq:binwidth}.
\ee
At BAO scales, this suggests that the minimal radial width which one
can bin data is therefore $\Delta \ln R \approx 10^{-6}$.  This
corresponds to exceedingly small angular bins, where the ML estimator
is expected to be much superior to the LS estimator.  In practice,
binning as fine as this is unnecessary, but it does highlight that the
ML estimator should enable finer binning than the LS estimator.

\acknowledgments We are grateful to Matthew Becker for illuminating
discussion and assistance regarding pair counting algorithms and for
providing an early version of the simulated galaxy catalogs.  We would
also like to thank the LasDamas team for making their numerical
simulations available.  This work was supported in part by the
U.S. Department of Energy contract to SLAC no. DE-AC02-76SF00515, and
by NASA through the Einstein Fellowship Program, grant PF9-00068.
Support was also provided by the Kavli Institute for Cosmological
Physics at the University of Chicago through grant NSF PHY-1125897,
and an endowment from the Kavli Foundation and its founder Fred Kavli.


\bibliography{v_apj}

\newcommand\AAA[3]{{A\& A} {\bf #1}, #2 (#3)}
\newcommand\PhysRep[3]{{Physics Reports} {\bf #1}, #2 (#3)}
\newcommand\ApJ[3]{ {ApJ} {\bf #1}, #2 (#3) }
\newcommand\PhysRevD[3]{ {Phys. Rev. D} {\bf #1}, #2 (#3) }
\newcommand\PhysRevLet[3]{ {Physics Review Letters} {\bf #1}, #2 (#3) }
\newcommand\MDRAS[3]{{MDRAS} {\bf #1}, #2 (#3)}
\newcommand\PhysLet[3]{{Physics Letters} {\bf B#1}, #2 (#3)}
\newcommand\AJ[3]{ {AJ} {\bf #1}, #2 (#3) }
\newcommand\aph{astro-ph/}
\newcommand\AREVAA[3]{{Ann. Rev. A.\& A.} {\bf #1}, #2 (#3)}

\newpage
\clearpage
\newpage

\appendix

\section{Derivation of $\Var(DD)$}
\label{app:varDD}

We present here a derivation of our expression for $\Var(DD)$ in
Eqs. \ref{eq:cov_poisson} and \ref{eq:Cmat_clustering}.  The remaining
terms in the covariance matrix can be derived in a similar fashion.  We have by definition
\begin{eqnarray}
\Var(DD) = \left< \left(DD(r)\right)^2 \right> - \left<
DD(r)\right>^2.
\end{eqnarray}
It was shown in the text that
\begin{eqnarray}
\left< DD(r) \right> = \frac{1}{2} \ndata^2 V V_1 \left[ 1 + \xi(r) \right].
\end{eqnarray}
Considering the remaining term in $\Var (DD)$ and using Eqs. \ref{eq:Di} and \ref{eq:dd}, we find
\begin{eqnarray}
\left< (DD(r))^2 \right> &=& \frac{1}{4} \left<\sum_{ijkl} \Delta V^4 \ndata^4 (1 + \delta_i)(1+\delta_j)(1+\delta_k)(1+ \delta_l) W_{ij} W_{kl} S_i S_j S_k S_l\right>\\
&=& \frac{1}{4} \left<\sum_{ijkl} \Delta V^4 \ndata^4 (1 + \delta_i\delta_j + \delta_i \delta_k + \delta_i \delta_l + \delta_j \delta_k + \delta_j \delta_l + \delta_k \delta_l + \right. \nonumber \\
&& \left. \delta_i \delta_j \delta_k + \delta_i \delta_j \delta_l + \delta_i \delta_k \delta_l + \delta_j \delta_k \delta_l + \delta_i\delta_j\delta_k\delta_l) W_{ij} W_{kl} S_i S_j S_k S_l   \vphantom{\sum_{ijkl}} \right>\\
&=& \frac{1}{4} \sum_{ijkl} \Delta V^4 \ndata^4 \left[1 + \frac{\delta_{ik}}{\ndata \Delta V} + \frac{\delta_{il}}{\ndata \Delta V} + \frac{\delta_{jk}}{\ndata \Delta V} + \frac{\delta_{jl}}{\ndata \Delta V} +  \right. \xi_{ij} + \xi_{ik} + \xi_{il} + \xi_{jk} + \xi_{jl} + \xi_{kl} \nonumber \\
&&  \left. + \left<\delta_i \delta_j \delta_k + \delta_i \delta_j \delta_l + \delta_i \delta_k \delta_l + \delta_j \delta_k \delta_l + \delta_i\delta_j\delta_k\delta_l \right> \vphantom{\frac{\delta_{jl}}{\ndata \Delta V}} \right] W_{ij} W_{kl} S_i S_j S_k S_l \vphantom{\sum_{ijkl}}\\
&=& \frac{1}{4}\sum_{ijkl} \Delta V^4 \ndata^4 W_{ij} W_{kl} S_i S_j S_k S_l   + \sum_{ij} \Delta V^3 \ndata^3  W_{ij} W_{ik} S_i S_j S_k + \frac{1}{2}\sum_{ij} \Delta V^4 \ndata^4 \xi(r) W_{ij}W_{kl} S_i S_j S_k S_l  + \ndata^4 \sum_{ijkl} \Delta V^4 \xi_{ik} W_{ij} W_{kl} S_i S_j S_k S_l  \nonumber \\
&& + \ndata^4 \left<\sum_{ijkl} \Delta V^4  \delta_i \delta_j \delta_k W_{ij}W_{kl} S_i S_j S_k S_l\right> + \frac{1}{4}\ndata^4 \left<\sum_{ijkl} \Delta V^4 \delta_i \delta_j \delta_k \delta_l W_{ij}W_{kl} S_i S_j S_k S_l\right> \\
&=& \frac{1}{4} (\ndata^2 VV_1)^2 + \ndata^3 VV_1V_2 + \frac{1}{2} \xi(r) (\ndata^2 VV_1)^2 + \ndata^4 \sum_{ijkl} \Delta V^4 \xi_{ik} W_{ij} W_{kl} S_i S_j S_k S_l  \nonumber \\
&& + \ndata^4 \left<\sum_{ijkl} \Delta V^4  \delta_i \delta_j \delta_k W_{ij}W_{kl} S_i S_j S_k S_l\right> + \frac{1}{4}\ndata^4 \left<\sum_{ijkl} \Delta V^4 \delta_i \delta_j \delta_k \delta_l W_{ij}W_{kl} S_i S_j S_k S_l\right> 
\end{eqnarray}
Substituting back into the expression for $\Var(DD)$ we have
\begin{eqnarray}
\Var(DD) &=& \ndata^3 VV_1V_2 + \ndata^4 \sum_{ijkl} \Delta V^4 \xi_{ik} W_{ij} W_{kl} S_i S_j S_k S_l + \ndata^4 \left<\sum_{ijkl} \Delta V^4  \delta_i \delta_j \delta_k W_{ij}W_{kl} S_i S_j S_k S_l\right> + \nonumber \\
&& \frac{1}{4}\ndata^4 \left<\sum_{ijkl} \Delta V^4 \delta_i \delta_j \delta_k \delta_l W_{ij}W_{kl} S_i S_j S_k S_l\right>  - \frac{1}{4} (\ndata^2 VV_1 \xi(r))^2.
\label{eq:dd_expanded}
\end{eqnarray}
The second term in the above expression can be re-written as
\begin{eqnarray}
\sum_{ijkl} \Delta V^4  \xi_{ik} W_{ij}W_{kl}S_iS_jS_kS_l &=& \sum_{ik} \Delta V^2  \xi_{ik} S_i S_k \sum_j \Delta V W_{ij} S_j \sum_l \Delta V W_{kl} S_l \\
&=& V_{1}^2 \sum_{ij} \Delta V^2  \xi_{ij}S_i S_j \\
&=& V_1^2 V^2 \twopt,
\end{eqnarray}
where we have used the definition of $\twopt$ in Eq. \ref{eq:twopt}.  The third term on the right hand side of Eq. \ref{eq:dd_expanded} can be written as
\begin{eqnarray}
\left<\sum_{ijkl} \Delta V^4  \delta_i \delta_j \delta_k W_{ij}W_{kl} S_i S_j S_k S_l\right> &=&  \left<\sum_{ijk} \Delta V^3  \delta_i \delta_j \delta_k W_{ij} S_i S_j S_k \sum_l \Delta V W_{kl} S_l \right> \\
&=& V_{1} \left<\sum_{ijk} \Delta V^3  \delta_i \delta_j \delta_k W_{ij} S_i S_j S_k\right>  \\
&=& V_1 (V_1 V^2) \threept,
\end{eqnarray}
where we have used the definition of $\threept$ in Eq. \ref{eq:threept}.  Finally,
we do a cumulant expansion of the fourth order term to separate out the
Gaussian contribution.  We have
\begin{eqnarray}
\frac{1}{4}\ndata^4 \left<\sum_{ijkl} \Delta V^4 \delta_i \delta_j \delta_k \delta_l W_{ij}W_{kl} S_i S_j S_k S_l\right> &=& \frac{1}{4}\ndata^4 \sum_{ijkl} \Delta V^4 \left[ \left<\delta_i \delta_j \right> \left<\delta_k \delta_l \right> + \left<\delta_i \delta_k\right> \left<\delta_j \delta_l \right> + \left<\delta_i \delta_l \right> \left<\delta_j \delta_k \right> + C_{ijkl}^{(4)}   \right] W_{ij}W_{kl} S_i S_j S_k S_l \\
&=& \frac{1}{4}\ndata^4 \sum_{ijkl} \Delta V^4 \left[ \xi_{ij}\xi_{kl} + 2\xi_{ik}\xi_{jl} + 2\frac{\delta_{ik}\delta_{jl}}{\ndata^2 \Delta V^2}  + C_{ijkl}^{(4)}   \right] W_{ij}W_{kl} S_i S_j S_k S_l \\
&=& \frac{1}{4} \ndata^4 (VV_1\xi(r))^2 + \frac{1}{2}\ndata^4 \sum_{ijkl} \Delta V^4 \xi_{ik}\xi_{jl}  W_{ij}W_{kl} S_i S_j S_k S_l + \nonumber \\
&& \frac{1}{2}\ndata^2 VV_1 +  \frac{1}{4}\ndata^4 \sum_{ijkl} \Delta V^4 C_{ijkl}^{(4)} W_{ij}W_{kl} S_i S_j S_k S_l \\
&=& \frac{1}{4} \ndata^4 (VV_1\xi(r))^2 + \frac{1}{2}\ndata^4 (V_1 V)^2 \fourpt + \frac{1}{2}\ndata^2 VV_1,
\end{eqnarray}
where we have used the definition of $\fourpt$ in Eq. \ref{eq:fourpt}.

Substituting the above results into our expression for $\Var(DD)$ we
have
\begin{eqnarray}
\Var(DD) =  \left[ \ndata^3 VV_1V_2  + \frac{1}{2}\ndata^2 VV_1 \right]  + \left[ \ndata^4 (V_1^2 V^2) \twopt + \ndata^4 (V_1^2 V^2) \threept + \frac{1}{2} \ndata^4 (V_1^2 V^2) \fourpt \right],
\end{eqnarray}
in agreement with Eqs. \ref{eq:cov_poisson} and
\ref{eq:Cmat_clustering}.


\clearpage
\newpage
\clearpage

\end{document}